\documentclass[usenatbib]{mn2e}
\usepackage{graphicx}
\usepackage{epstopdf}
\usepackage{natbib}
\usepackage{verbatim}
\usepackage{amssymb,amsmath}
\usepackage{ulem}
\usepackage{color}
\usepackage[bookmarks,bookmarksnumbered,colorlinks=true, citecolor=blue, linkcolor=black]{hyperref}
\usepackage[mathscr]{eucal}

\usepackage[usenames,dvipsnames]{xcolor}  

\pdfoutput=1

\def \ltsim {\lesssim}

\renewcommand{\thefootnote}{\fnsymbol{footnote}}

\newcommand\apj{ApJ}
\newcommand\apjl{ApJ}
\newcommand\apjs{ApJS}
\newcommand\aap{A\&A}
\newcommand\mnras{MNRAS}
\newcommand\nat{Nature}

\title[Number Density Evolution]{An analysis of the evolving comoving number density of galaxies in hydrodynamical simulations}

\author[P. Torrey et al.]
       {\parbox{18cm}{Paul~Torrey$^{1,2}$\footnotemark[1], Sarah Wellons$^{3}$, Francisco Machado$^{1}$, Brendan Griffen$^{1}$,\\ Dylan Nelson$^{3}$,
       Vicente Rodriguez-Gomez$^{3}$, Ryan McKinnon$^{1}$, Annalisa Pillepich$^{3}$, Chung-Pei Ma$^{4}$,  Mark Vogelsberger$^{1}$, Volker Springel$^{5,6}$ and Lars Hernquist$^{3}$
       }\vspace{0.3cm}\\ 
         $^1 $ MIT Kavli Institute for Astrophysics \& Space Research, Cambridge, MA 02139, USA\\
         $^2 $ TAPIR, Mailcode 350-17, California Institute of Technology, Pasadena, CA 91125, USA \\
         $^3$ Harvard-Smithsonian Center for Astrophysics, 60 Garden Street, Cambridge, MA 02138, USA\\ 
         $^4$ Astronomy Department, University of California at Berkeley, Berkeley, CA 94720, USA\\
         $^5$ Heidelberger Institut f\"{u}r Theoretische Studien, Schloss-Wolfsbrunnenweg 35, 69118 Heidelberg, Germany\\
         $^6$ Zentrum f\"{u}r Astronomie der Universit\"{a}t Heidelberg, ARI, M\"onchhofstr. 12-14, 69120 Heidelberg, Germany\\
         }

\begin{document}

\maketitle

\begin{abstract}
The cumulative comoving number-density of galaxies as a function of stellar mass
or central velocity dispersion is commonly used to link galaxy populations across
different epochs. 
By assuming that galaxies preserve their number-density in time,
one can infer the evolution of their properties, such as masses, sizes, and
morphologies. 
However, this assumption does not hold in the presence of galaxy
mergers or when rank ordering is broken owing to variable stellar growth rates. 
We present an analysis of the evolving comoving number density of galaxy populations found in the Illustris cosmological hydrodynamical simulation focused on the redshift range $0\leq z \leq 3$.
Our primary results are as follows: 
1) The inferred average stellar mass evolution obtained via a constant comoving number density assumption is systematically biased compared to the merger tree results at the factor of $\sim$2(4) level when tracking galaxies from redshift $z=0$ out to redshift $z=2(3)$;
2) The median number density evolution for galaxy populations tracked forward in time is shallower than for galaxy populations tracked backward in time;
3) A similar evolution in the median number density of tracked galaxy populations is found regardless of whether number density is assigned via stellar mass, stellar velocity dispersion, or dark matter halo mass;
4) Explicit tracking reveals a large diversity in galaxies' assembly histories that cannot be captured by constant number-density analyses;
5) The significant scatter in galaxy linking methods is only marginally reduced by considering a number of additional physical and observable galaxy properties as realized in our simulation.
We provide fits for the forward and backward median evolution in stellar mass and number density for use with observational data and
discuss the implications of our analysis for interpreting multi-epoch galaxy property observations as related to galaxy evolution.
\end{abstract}


\renewcommand{\thefootnote}{\fnsymbol{footnote}}
\footnotetext[1]{E-mail: ptorrey@mit.edu}

\section{Introduction}
Modeling galaxy evolution based on observational data requires a method for linking galaxy populations between different epochs.  
Establishing direct progenitor-descendant links would allow for reconstruction of the mass, size, star formation rate, color, and morphology evolution (among other things) of galaxies directly from observational data.
However, formulating a method that accurately links progenitor and descendant galaxy populations is non-trivial.
Incorrectly linking observed galaxy populations between different epochs results in errors in the inferred evolutionary tracks.
This effect has come to be known as ``progenitor bias''~\citep[e.g.,][]{vD1996, Saglia2010} and is a well-known effect that needs to be addressed in order to infer the mass~\citep{vanDokkum2010, Brammer2011, Patel2013, vanDokkum2013}, size~\citep{Fan2008, Valentinuzzi2010, Carollo2013, Patel2013, vanDokkum2013, Morishita2015}, star formation rate~\citep{Papovich2011} and morphology evolution~\citep{vanDokkum2001, Daddi2005}.

Several approximations have been employed to estimate progenitor-descendant linking of galaxy populations in order to minimize or remove progenitor bias.
One method that has been applied is to select rare or distinct galaxy populations which one might reasonably be able to recover at different observational epochs.
Such analysis has been applied to examine the redshift-dependent properties of brightest cluster galaxies~\citep[BCGs; e.g.,][]{Butcher1984, Aragon1993, Lidman2012, Vulcani2014} 
as well as to populations of massive early-type galaxies~\citep[ETGs; e.g.,][]{Daddi2005, Trujillo2007, Barro2013}.
The premise behind this linking metric is that both massive ETGs and BCGs observed at high redshift will remain massive ETGs and BCGs into the low-redshift universe.
This assumption is generally true.
However, caution must be taken as the fraction of galaxies that are massive and quenched grows with time.
This will drive an increase in the number of massive ETGs and BCGs that exist in the local universe compared to the high-redshift universe, and therefore create contamination in the inferred progenitor-descendant galaxy populations.
It has been argued in~\citet{Carollo2013} that this effect alone can result in the inferred size evolution of massive ETGs~\citep[but see also][for careful analysis that indicates progenitor bias is insufficient to fully explain the observed size evolution]{Belli2014, Keating2015}.
The current uncertainty that surrounds the size evolution of ETGs is dominated by a lack of theoretical understanding of how to link high and low redshift galaxy populations.

Another approach -- which is the focus of this paper -- is to assume that progenitor and descendant galaxy populations can be linked based on their cumulative comoving number density~\citep[e.g.,][]{Wake2006, vanDokkum2010, Papovich2011, Brammer2011, Patel2013, vanDokkum2013}.
This method is appealing because it provides a straightforward way to infer information about galaxy evolution directly from multi-epoch cumulative stellar mass functions.
Because no assumptions about galactic characteristics (e.g., massive and quenched) are required, this method has been employed to infer the mass evolution of Milky Way progenitors~\citep{Patel2013, vanDokkum2013}, 
the evolution of red/quenched galaxy fractions~\citep{Brammer2011},
and the size and morphological evolution of massive galaxies~\citep{vanDokkum2010}.

Linking galaxy populations in this fashion implicitly involves two assumptions: (i) that the total number (density) of galaxies is conserved and (ii) that galaxies maintain their rank order.
If both of these assumptions are true, then the most(least) massive galaxies at some initial redshift (e.g., $z=2$) will still be the most(least) massive galaxies at any other redshift (e.g., $z=0$).
The two primary issues with this approach are that galaxies undergo merger events which can reduce their number density, and that individual galaxies may have scattered/stochastic growth histories which can break their rank ordering.
Mergers will drive an evolution in the number density of any given galaxy population by simply changing the number density of the total galaxy population with time, even if galaxies remain well rank-ordered.
On the other hand, scattered growth rates will cause a population of galaxies with similar initial stellar mass at some initial redshift (e.g., $z=2$) to have some potentially wide distribution of masses by the time they evolve to some new observational epoch (e.g., $z=0$).
To avoid unnecessary biases when applying constant comoving number-density analysis to observational datasets, it is important to understand the extent to which galaxy mergers or variable galaxy growth rates impact number-density matching.

Capturing galaxy mass assembly, including both mergers and variable mass growth rates, can be done in detail by employing numerical galaxy formation simulations.
The efficiency of constant comoving number-density selections has been studied using semi-analytic models~\citep{Leja2013, Mundy2015} and with abundance matching~\citep{Behroozi2013}.  
\citet{Leja2013} employed the~\citet{Guo2011} semi-analytic models based on the Millennium Simulation~\citep{Millennium} to compare tracked mass evolution against assumed mass evolution based on number-density selections.
They found that a constant comoving number-density selection yielded inferred median descendant masses of high-redshift galaxy populations which differed by 40\% from the actual descendant masses.
By applying a correction to account for the scatter in galaxy growth rates and mergers they were able to reduce the mass offset error from a number-density selection to 12\%.
However, even with such a correction, significant scatter remains among the inferred growth rates.

\citet{Mundy2015} used several different semi-analytic models also based on the Millennium Simulation~\citep{Millennium} to provide fitting functions that describe the ``recovery fraction" of galaxies with time.
They found that for their lowest number-density bin (corresponding to the most massive/rare haloes) a constant number-density selection recovered roughly $30\%$ of available descendants, with the majority of the recovered galaxy population being contamination.
\citet{Behroozi2013} used an abundance matching based on the Bolshoi Simulation~\citep{Bolshoi} to infer the number-density evolution of galaxies.
They found a median evolution in the number density for all galaxy mass bins, and provided fitting functions for the median evolution.
These fitting functions have been applied in~\citet{Marchesini2014}, where the mass and star formation rate histories of massive galaxies were inferred using the prescribed number density evolution tracks.

Our work builds on these previous theoretical studies by examining the constant comoving number-density selection method using a full volume hydrodynamical simulation.
Multi-epoch constant comoving number-density selection method has only recently been examined using hydrodynamical simulations at high redshifts~\citep{Jaacks2015}, with limited study at $z<3$.
To the extent that both hydrodynamical simulations and semi-analytic methods accurately reproduce the evolution of the galaxy stellar mass function~\citep[see][for a comparison and discussion]{Somerville2014}, it is unlikely that full hydrodynamical simulations would yield very different results compared to their semi-analytic model counterparts.
However, there are some subtle but important issues that we might expect to yield concrete differences between the previous analyses and what we present in this paper.

Specifically, the cosmology used in the original Millennium Simulation~\citep{Millennium} was based on the WMAP-1 data release~\citep{WMAPI}, and therefore applied a $\sigma_8$ value that is higher than what is currently accepted.
This has the consequence of boosting the halo-halo merger rate, which is of direct relevance to galaxy number-density analysis~\citep{Leja2013, Mundy2015}.  
A direct comparison of the halo-halo merger rates between the Millennium Simulation and Illustris shows that the merger rates are in good agreement despite differences in the adopted cosmological parameters~\citep{RodriguezGomez2015}.
The galaxy-galaxy merger rates used in semi-analytic models can be significantly offset from the galaxy-galaxy merger rates found in Illustris~\citep{RodriguezGomez2015} which could impose differences on the number density evolution.
However, given that the results presented in~\citet{Behroozi2013} (which used updated cosmological parameters) generally agreed with~\citet{Leja2013}, it is unlikely that updated cosmological parameters alone would drive major differences in the number-density evolution of galaxies.

Perhaps more importantly, while semi-analytic models and hydrodynamical simulations both attempt to include a similar array of physical processes (gas cooling, star formation, feedback, etc.), in detail the growth rates of galaxies are calculated in very different ways.
It is therefore not guaranteed that the scattered growth rates that can contribute to breaking galaxy rank order will be equally realized in semi-analytic models and hydrodynamical simulations.
Moreover, while semi-analytic models are able to infer a wide range of internal galactic properties (e.g., galaxy size, bulge-to-disk ratio, etc.), all of these properties are managed and evolved at a sub-grid level.
Although the hydrodynamical simulations employed in this paper also apply sub-grid models to manage many aspects of galaxy formation physics (e.g., star formation, ISM gas phase structure, etc.) some important galactic characteristics such as stellar velocity dispersion can be self-consistently evolved.
Finally, the general analysis presented in this paper complements nicely the detailed studies of specific galaxy populations that form in our simulations and have been compared against observations, such as the formation and evolution of compact massive galaxies~\citep{Wellons2015a, Wellons2015b}.

The paper is structured as follows: 
In Section~\ref{sec:Methods} we discuss our methods including a description of the simulations that have been employed, the construction of galaxy property catalogs, and the merger trees that form the core of our analysis.
In Section~\ref{sec:Results_CMF} we present the simulated multi-epoch cumulative stellar mass function and the corresponding inferred stellar mass evolution.   
We compare this against the actual tracked stellar mass evolution found in the simulation.  We introduce useful fitting formulae here, which describe the mass and number-density evolution for the explicitly tracked galaxy population.
In Section~\ref{sec:Results_VDF} we present the simulated multi-epoch cumulative velocity dispersion function and consider whether velocity dispersion might act as a better galactic parameter for linking galaxy populations.
In Section~\ref{sec:Discussion} we discuss our results, with a focus on understanding the empirical origin of the trends we find in our simulations and exploring implications for using number-density analyses on observational data sets.
We conclude in Section~\ref{sec:Conclusions}.

\section{Methods}
\label{sec:Methods}
In this paper we use the Illustris simulation to study the relationship between the stellar mass, dark matter mass, central stellar velocity dispersion, and number-density evolution of galaxies.  
Full details of the Illustris project can be found in~\citet{Vogelsberger2014b, Vogelsberger2014a, Genel2014}.

Briefly, the Illustris simulation is a cosmological hydrodynamical simulation run in a periodic box of size $L=106.5$ Mpc.
Illustris was run using the {\small AREPO} simulation code~\citep{AREPO} using a physical setup that
includes gravity, hydrodynamics, radiative cooling of gas~\citep{KatzCooling}, 
star formation with associated feedback~\citep{SH03}, 
mass and metal return to the interstellar medium from aging stellar populations~\citep{WiersmaGasReturn, Vogelsberger2013}, 
and supermassive black hole growth with associated feedback~\citep{Dimatteo2005, Springel2005a, Sijacki2007, Vogelsberger2013, Sijacki2014}.  
The feedback models employed in the simulation were chosen to match the redshift $z=0$ galaxy stellar mass function and cosmic star formation rate history, and it has been subsequently shown that
it broadly reproduces the observed evolving galaxy stellar mass function out to high redshift~\citep{Torrey2014, Genel2014, Sparre2014}.
The relatively large volume allows for sampling across a range of galaxy environments~\citep{Vogelsberger2014b} including rare objects~\citep[e.g., compact massive galaxies,][]{Wellons2015a} with diverse formation histories~\citep{Sparre2014}, all of which is important for the present work.
The Illustris simulation contains roughly $1820^3$ baryon and dark matter particles yielding a baryon mass resolution of $M_{{\rm bar} }\approx 1.3 \times10^6 M_\odot$ and a dark matter mass resolution of $M_{{\rm DM}} = 6.3 \times 10^{6} M_\odot$ (The number of dark matter particles remains exactly fixed at $1820^3$ for the whole run, but the number of baryonic resolution elements changes owing to cell (de)refinement).
The Plummer-equivalent gravitational force softening lengths used in the simulation is $\epsilon=1.0 \; h^{-1}$ ckpc for both dark matter and baryons until $z=1$, at which time the baryonic gravitational softening length is capped at a maximum physical value of $\epsilon=0.5 \;  h^{-1}$ pkpc.  (The dark matter gravitational softening length continues at a fixed comoving size to $z=0$.) 

Several steps have been taken to post-process the Illustris data output to facilitate the present analysis.
First, the simulation output is run through {\small SUBFIND} to identify friends-of-friends (FoF) haloes and bound sub-haloes~\citep{SUBFIND, Dolag2009}.
Throughout this paper, we employ the {\small SUBFIND} sub-halo catalog to identify galaxy populations, including both centrals and satellites.
Wherever we refer to galaxies or galaxy populations, we are in detail referring to the self-bound sub-halo structures identified by {\small SUBFIND}.

Second, a wide range of physical properties -- including stellar mass, star formation rate, half-mass radius, etc. -- of each structure identified with {\small SUBFIND} have been tabulated.
A catalog of galaxy properties is calculated for each galaxy and each redshift independently.  
Throughout this paper we use stellar masses and dark matter halo masses.
In both cases, we calculate the stellar (dark matter halo) masses as being the total mass of all gravitationally bound stellar (dark matter) particles of a given {\small SUBFIND} subhalo.
For this paper, we have additionally calculated the stellar velocity dispersion for the galaxy population defined as the three dimensional standard deviation of stellar particle velocities calculated within the stellar half-mass radius.

The third post-processing step is to link the galaxy catalogs together in time using merger trees.
In this paper we adopt the SubLink merger trees as described in~\citet{RodriguezGomez2015}.
The merger trees are constructed by identifying progenitor/descendant galaxy pairings based on overlapping particle compositions identified through particle identification numbers.
The merger trees facilitate tracking of individual galaxies forward and backward in time while including in situ growth and contributions from mergers.
When galaxy mergers occur, the progenitors are segregated into a single main branch and secondary progenitors.
We define the main progenitor branch as being the most massive branch when summed over the entire formation history until that point~\citep{DeLucia2007, RodriguezGomez2015}.
Other operational definitions of main progenitor branch are possible~\citep[e.g., most massive halo at the previous snapshot,][]{Millennium} and some of the results quoted in this paper depend on this assumption.
However, we have verified that this choice has a very limited impact on our results, with all of our results being qualitatively invariant to this choice.

The full data from the Illustris simulation -- including all data, post-processing {\small SUBFIND} galaxy property catalogs, merger trees data, and basic scripts and procedures required to reproduce our analysis -- have been made publicly available~\citep{Nelson2015}.\footnote{\url{http://www.illustris-project.org}}

\section{Results:  Tracing Galaxies via Stellar Mass}
\label{sec:Results_CMF}
\subsection{Cumulative Stellar Mass Function}
\label{sec:CMF}
Perhaps the most relevant aspect of our model for this paper is its ability to reproduce the (cumulative) galaxy stellar mass function at many observational epochs.
It has also been shown that the feedback model employed by Illustris -- described in detail in~\citet{Vogelsberger2013} -- is capable of producing a galaxy stellar mass function and star formation main sequence that broadly matches observations~\citep{Torrey2014, Genel2014, Sparre2014}.
This agreement is achieved through a combination of star formation driven winds to moderate star formation in low mass galaxies, and AGN feedback to regulate the growth of massive galaxies.
This combination of feedback results in a multi-epoch galaxy stellar mass function that is similar to modern semi-analytic models and other hydrodynamical simulations~\citep[see][for comprehensive review plots and discussion]{Somerville2014}.  
Here, we present fits to the redshift evolution of the cumulative galaxy stellar mass function as found in our simulations.
This fit is important to the analysis that we carry out in subsequent sections of the paper.

\begin{figure}
\centerline{\vbox{\hbox{
\includegraphics[width=3.5in]{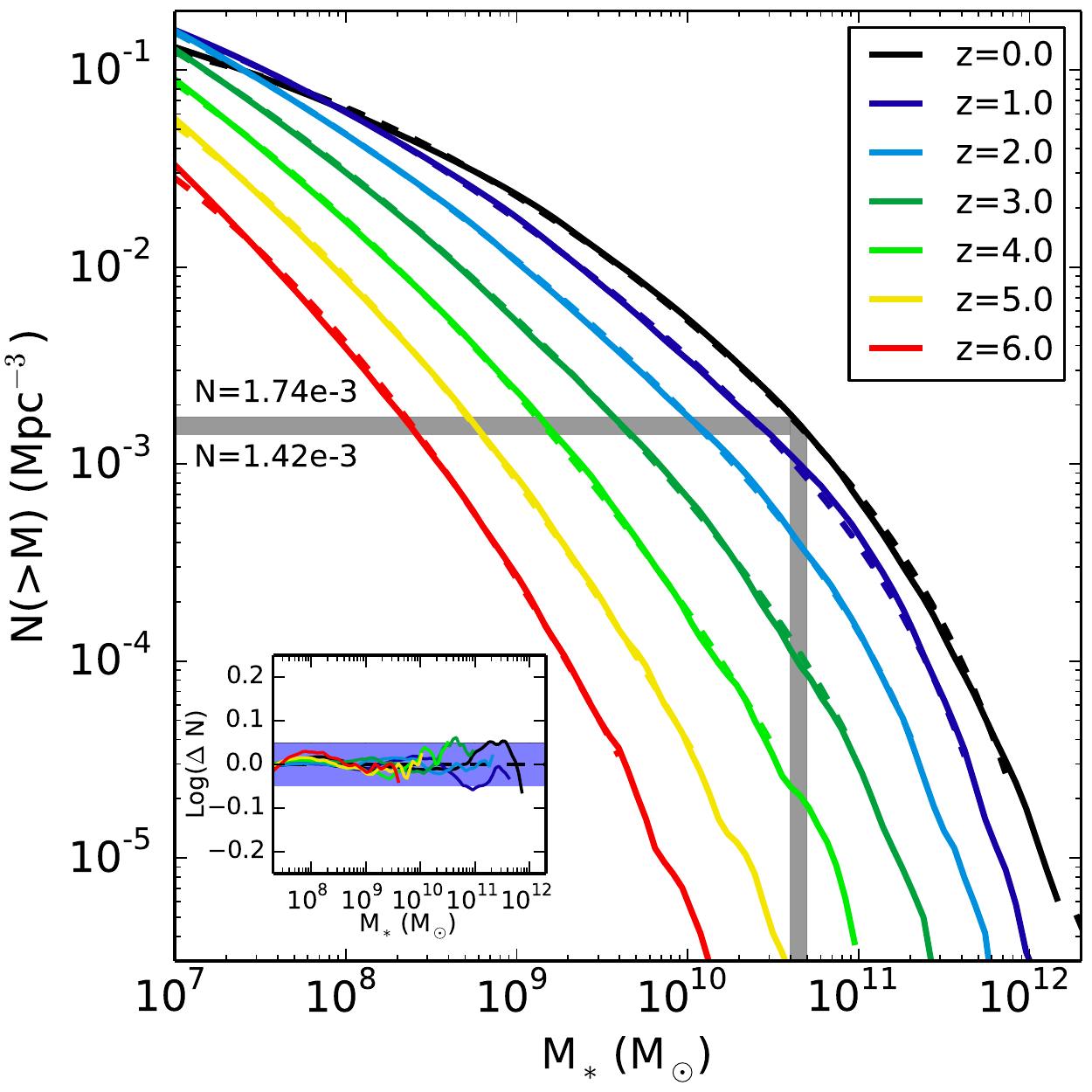}
}}}
\caption{Cumulative stellar mass functions derived from the galaxy populations found in Illustris are shown at several redshifts as indicated in the legend.  
The grey region identifies the stellar mass range (vertical strip) and cumulative number density range (horizontal strip) that correspond to the Milky Way mass objects at redshift $z=0$ as defined and discussed in the text.
The dashed lines shown within indicate the multi-epoch CMF fitting functions.
The fitting functions nearly overlap with the actual CMFs at all redshifts, and so we also show the ``error" associated with these fits in the panel inset, with the solid blue band indicating 5\% errors.
The mass evolution of galaxies can be inferred from the fitting functions by identifying the mass associated with a constant comoving number density at several redshifts (e.g., where the grey horizontal band intersects the CMFs).
 }
\label{fig:cum_number_density}
\end{figure}

Figure~\ref{fig:cum_number_density} shows the cumulative mass function (CMF) at several redshifts as realized in the Illustris simulation.
We fit the simulated cumulative galaxy stellar mass function with a power law plus exponential dependence of the form
\begin{equation}
N =  A \; \tilde{ M}_* ^{\alpha + \beta \mathrm{Log}\tilde{ M}_*  } \; {\rm exp} ( -\tilde{ M}_*  )
\label{eqn:CumMassFunc}
\end{equation}
where $\tilde{ M}_* =M_*/(10^{\gamma} M_\odot)$.  
The combined power-law, exponential form of Equation~(\ref{eqn:CumMassFunc}) is adopted to be similar to the ~\citet{SchechterFunction} function commonly used to describe galaxy stellar mass and luminosity functions.
We allow all of the fit variables to vary with redshift according to
\begin{eqnarray}
A            &=& a_0            + a_1 z             + a_2 z^2 \label{eqn:A}\\
\alpha    &=& \alpha_0     + \alpha_1 z     + \alpha_2 z^2 \label{eqn:B}\\
\beta      &=& \beta_0       + \beta_1 z      + \beta_2 z^2 \label{eqn:C}\\
\gamma &=& \gamma_0  + \gamma_1 z + \gamma_2 z^2 \label{eqn:E}
\end{eqnarray}
where $z$ is redshift.
Adopting this fitting form results in 12 independent coefficients (all variables on the RHS of equations~\ref{eqn:A}-\ref{eqn:E}) that are set using an ordinary least squares regression on the CMFs over the redshift range $z=0$ to $z=6$, mass range $M_*>10^7 M_\odot$, and number density range $N>3 \times 10^{-5} {\rm Mpc^{-3}}$.  
The resulting fits are shown in Figure~\ref{fig:cum_number_density} as dashed lines, which can be compared against the solid lines that trace the CMFs taken directly from the simulation.
In the inset of Figure~\ref{fig:cum_number_density} we also show ${\rm Log}_{10} \left( \Delta N \right) = {\rm Log}_{10} \left( N_{{\rm sim}}/N_{{\rm fit}} \right)$ which gives an impression for the level of fit accuracy.
Using the power law plus exponential fit described above we obtain a fit to the CMF that is valid from $z=0$ to $z=6$ with typical errors of order $1\%$, which always remain well below $10\%$.
The best fit coefficients for the CMF in Illustris in the redshift range $0<z<6$ are given in Table~\ref{table:CMF_fit}.
The appropriate limits on this fitting function cover the mass range $10^7 M_\odot < M_* < 10^{12} M_\odot$, CMF values $\phi > 3 \times 10^{-5} \mathrm{Mpc}^{-3}$, and redshift range $0<z<6$. 
However, one should additionally bear in mind that the baryon particle mass in our simulations is $\sim10^6 M_\odot$, and so caution should be taken when considering the low-mass end of the mass function where only $10-100$ stellar particles are included in each galaxy.

Despite its 12 free terms, Equation~(\ref{eqn:CumMassFunc}) is trivial to calculate.
Upon evaluation, one can easily identify the cumulative number density of galaxies over the full resolved mass range in Illustris from $0<z<6$.
This is generally useful, including for comparisons with observed cumulative stellar mass functions -- which is outside the scope of this paper.\footnote{The CMF is most valuable to the present paper where we focus on the evolution of galaxies in number density space.  We additionally provide fits of the same form to the differential stellar mass function in Appendix~\ref{sec:DiffMassFunction} which is the more commonly adopted form for examining the galaxy stellar mass function.}
The general form allows us to obtain useful fits to a wide variety of smoothly varying functions making it possible to use expressions that take a similar form to Equation~(\ref{eqn:CumMassFunc}) at several points in this paper including the cumulative velocity dispersion function and tracked galaxy number-density evolution.
We are additionally making available with this paper simple python scripts that allow one to evaluate Equation~(\ref{eqn:CumMassFunc}).\footnote{\url{https://github.com/ptorrey/torrey_cmf}}

\begin{table}

\begin{center}
\caption{ Best-fit parameters to the redshift-dependent CMF presented in Equation~(\ref{eqn:CumMassFunc}).}
\label{table:CMF_fit}
\begin{tabular}{ c  c c c c c  }
\hline
  &      \multicolumn{1}{|c|}{$i=0$}     &       \multicolumn{1}{|c|}{$i=1$}      &     \multicolumn{1}{|c|}{$i=2$}     \\
\hline 
\hline 
$a_i$ & -2.893811  & 0.082199  & -0.123157   \\
$\alpha_i$ & -0.625598  & 0.086216  & -0.049033   \\
$\beta_i$ & -0.038895  & 0.025419  & -0.007130   \\
$\gamma_i$ & 11.523852  & -0.187102  & 0.021022   \\
\hline
\hline
\end{tabular}
\end{center}

\begin{center}
\caption{ Best-fit parameters to the backward-tracked number-density evolution.  The below parameters along with Equation~(\ref{eqn:CumMassFunc}) describe the median number density of a redshift $z=0$ selected galaxy population at some higher redshift out to $z=3$.}
\label{table:Zevo_fit}
\begin{tabular}{ c  c c c c c  }
\hline
  &      \multicolumn{1}{|c|}{$i=0$}     &       \multicolumn{1}{|c|}{$i=1$}      &     \multicolumn{1}{|c|}{$i=2$}     \\
\hline 
\hline 
$a_i$ & -2.643961  & -0.299579  & -0.037861   \\
$\alpha_i$ & -0.526844  & -0.138136  & -0.032246   \\
$\beta_i$ & -0.026482  & -0.016006  & -0.005645   \\
$\gamma_i$ & 11.339278  & 0.391025  & -0.015732   \\
\hline
\hline
\end{tabular}
\end{center}

\begin{center}
\caption{Best-fit parameters to the forward-tracked number-density evolution starting from $z=1$.  The fit is only valid from $z=1$ to $z=0$.}
\label{table:fwd1}
\begin{tabular}{ c  c c c c c  }
\hline
  &      \multicolumn{1}{|c|}{$i=0$}     &       \multicolumn{1}{|c|}{$i=1$}      &     \multicolumn{1}{|c|}{$i=2$}     \\
\hline 
\hline 
$a_i$ & -3.640099  & 1.036010  & -0.311620   \\
$\alpha_i$ & -0.797215  & 0.292991  & -0.077466   \\
$\beta_i$ & -0.048383  & 0.033631  & -0.004868   \\
$\gamma_i$ & 11.827591  & -0.665051  & 0.184837   \\
\hline
\hline
\end{tabular}
\end{center}

\begin{center}
\caption{Best-fit parameters to the forward-tracked number-density evolution starting from $z=2$.  The fit is only valid from $z=2$ to $z=0$.}
\label{table:fwd2}
\begin{tabular}{ c  c c c c c  }
\hline
  &      \multicolumn{1}{|c|}{$i=0$}     &       \multicolumn{1}{|c|}{$i=1$}      &     \multicolumn{1}{|c|}{$i=2$}     \\
\hline 
\hline 
$a_i$ & -3.991685  & 0.348706  & 0.050966   \\
$\alpha_i$ & -0.793825  & 0.019329  & 0.036924   \\
$\beta_i$ & -0.038297  & -0.000232  & 0.006948   \\
$\gamma_i$ & 11.828429  & -0.280904  & -0.030598   \\
\hline
\hline
\end{tabular}
\end{center}

\begin{center}
\caption{ Best-fit parameters to the forward-tracked number-density evolution starting from $z=3$.  The fit is only valid from $z=3$ to $z=0$.}
\label{table:fwd3}
\begin{tabular}{ c  c c c c c  }
\hline
  &      \multicolumn{1}{|c|}{$i=0$}     &       \multicolumn{1}{|c|}{$i=1$}      &     \multicolumn{1}{|c|}{$i=2$}     \\
\hline 
\hline 
$a_i$ & -4.353012  & -0.438934  & 0.229633   \\
$\alpha_i$ & -0.792804  & -0.191351  & 0.066223   \\
$\beta_i$ & -0.029089  & -0.020006  & 0.007395   \\
$\gamma_i$ & 11.801074  & 0.197901  & -0.149748   \\
\hline
\hline
\end{tabular}
\end{center}

\begin{center}
\caption{ Best-fit parameters to the redshift-dependent cumulative velocity dispersion function presented in Equation~(\ref{eqn:CMVDFunc}).}
\label{table:CVDF_fit}
\begin{tabular}{ c  c c c c c  }
\hline
  &      \multicolumn{1}{|c|}{$i=0$}     &       \multicolumn{1}{|c|}{$i=1$}      &     \multicolumn{1}{|c|}{$i=2$}     \\
\hline 
\hline 
$a_i$ & 7.391498  & 5.729400  & -1.120552   \\
$\alpha_i$ & -6.863393  & -5.273271  & 1.104114   \\
$\beta_i$ & 2.852083  & 1.255696  & -0.286638   \\
$\gamma_i$ & 0.067032  & -0.048683  & 0.007648   \\
\hline
\hline
\end{tabular}
\end{center}

\end{table}

\subsection{Milky Way mass galaxies:  constant vs. non-constant number density }

We begin our analysis of the evolutionary tracks and evolving comoving number-densities of Illustris galaxies by considering the formation history of a population of Milky Way mass galaxies. 

We adopt a definition for ``Milky Way mass galaxies" as those galaxies with a redshift $z=0$ stellar mass in the range $4\times 10^{10}M_\odot < M_* < 5\times 10^{10}M_\odot$~\citep[e.g.,][]{McMillan2011, Bovy2013}.
This corresponds to $\sim$410 galaxies at redshift $z=0$, including all morphological types, formation histories, environments, etc. sampled in the simulation volume.
The vertical grey shaded region in Figure~\ref{fig:cum_number_density} indicates the redshift $z=0$ mass range adopted for Milky Way type galaxies in this section.
The corresponding horizontal grey shaded region identifies the cumulative number-density range that is associated with the redshift $z=0$ Milky Way galaxy mass range.
We note that in what follows, our results are not very sensitive to this specific choice of initial mass or number-density range.

\subsubsection{Examples of Individual Galaxy Evolutionary Tracks}
\begin{figure*}
\centerline{\vbox{\hbox{
\includegraphics[width=7.0in]{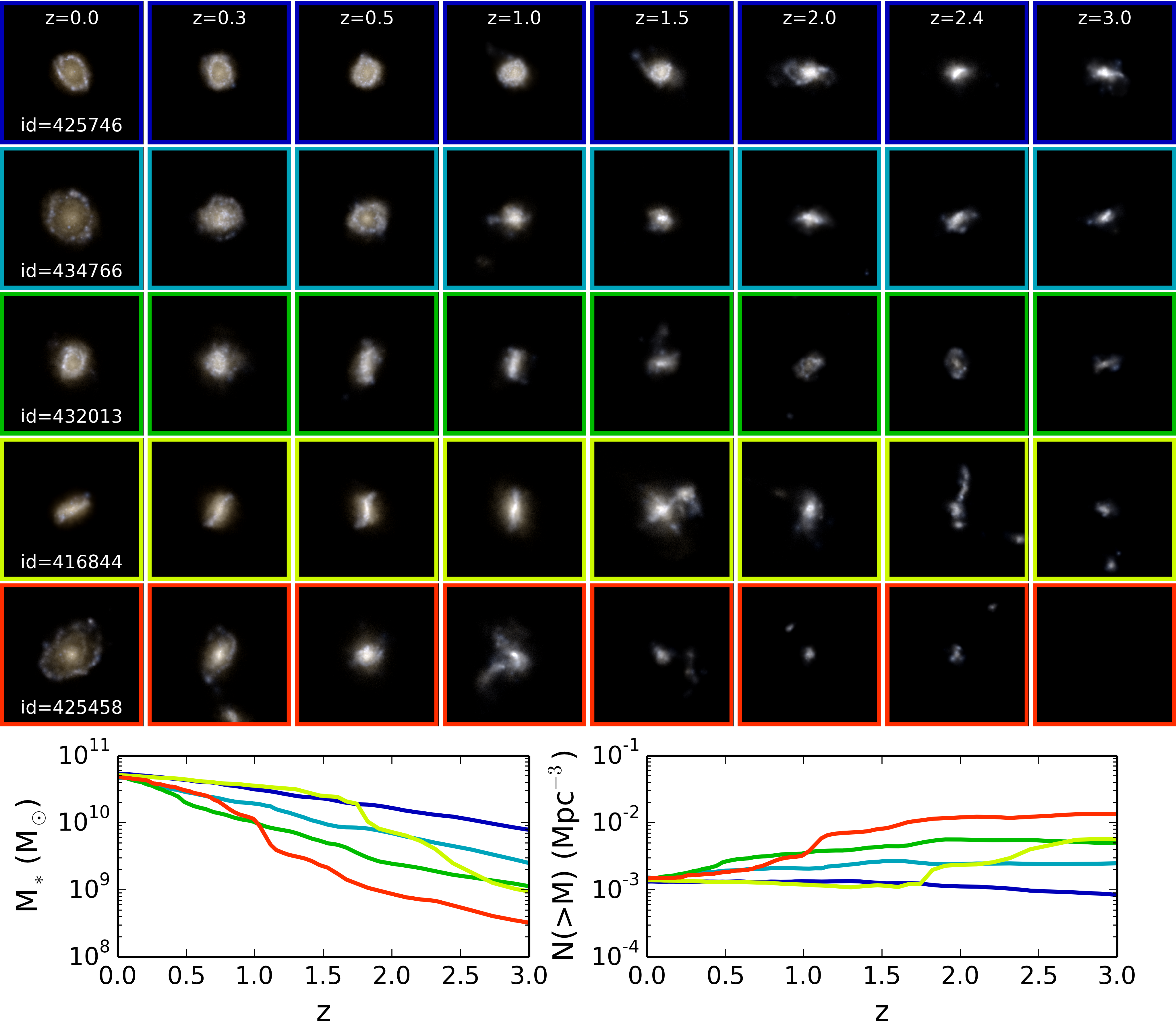}
}}}
\caption{
Mock stellar light images are shown for 5 galaxies taken from the Milky Way mass galaxy sample as described in the main text.  
Each system is shown at 8 redshifts from $z=0$ to $z=3$ to highlight the variety of formation histories that exist for galaxies with similar initial stellar masses.
The rows are ordered by increasing $z=3$ progenitor mass.
In some cases (the bottom two rows in particular) we find merger events contribute significantly to the growth of these systems. 
The bottom two panels show the mass (left) and cumulative number density (right) evolution for the 5 selected galaxies, with the color of the line corresponding to the border color for each image.  }
\label{fig:mw_morph_evo}
\end{figure*}

Figure~\ref{fig:mw_morph_evo} shows synthetic stellar light images~\citep{Torrey2015, Snyder2015} in the SDSS-g, -r, -i bands made for a selection of 5 galaxies from the Milky Way mass sample at 8 redshifts during their formation (as labeled within the top panel of each column). 
The progenitors of these 5 systems have been determined directly from the galaxy merger tree which finds the progenitor(s) of any galaxy based on the particle ID composition of each galaxy. 
If multiple progenitor galaxies exist while tracing galaxies backward in time we always select the ``main progenitor", defined as the progenitor with the most massive history~\citep{DeLucia2007, RodriguezGomez2015}.
We have ordered the galaxies by their redshift $z=3$ progenitor masses, with individual mass and number density evolution tracks shown in the bottom panels.  

We find that three of the systems (the top three) evolve without significant influence from mergers, with relatively smooth mass growth (and number density tracks).
These three systems preserve their rank order but diverge in their stellar mass from each other with time.
By redshift $z=3$, the top (blue) and middle (green) galaxies have stellar masses that are different by an order of magnitude.
The bottom two examples were selected to highlight systems that undergo mergers and change their rank order significantly with time.
For example, the top (blue) and fourth (yellow) galaxies have nearly identical mass evolution tracks from redshift $z=0$ out to redshift $z=1.5$.
At that time the fourth system can be identified through the postage stamp images to be undergoing a significant merger event, after which the blue and yellow mass growth tracks diverge.  
Despite their nearly identical mass evolution out to redshift $1.5$, these systems are offset by roughly an order of magnitude in their stellar mass by redshift $z=3$.
A similar qualitative story holds for the bottom (red) system, which follows a relatively median mass growth track out to redshift $z=0.7$, but quickly becomes the least massive of these galaxy progenitors thereafter.

The mass growth tracks shown in the left bottom panel of Figure~\ref{fig:mw_morph_evo} directly translate via the CMF into the number density evolution tracks shown in the bottom right panel.
We find that the dark and light blue galaxies which were already massive systems at redshift $z=3$ and followed mild growth paths thereafter are close to remaining on constant comoving number density evolution tracks.  
In contrast, the red system which grew rapidly since redshift $z=3$ has an evolution in number density of nearly an order of magnitude.
Figure~\ref{fig:mw_morph_evo} highlights the diversity in individual growth paths that occur at a fixed $z=0$ stellar mass.  
Using the full sample of Milky Way mass galaxies in the simulation, we can further consider the median growth tracks and the dispersion about those tracks for this galaxy population.

\subsubsection{Population Evolutionary Tracks}
We next consider the mass evolution of the full Milky Way mass selected galaxy population using now two complementary methods.
First, armed solely with an evolving set of cumulative stellar mass functions, we can identify the galaxy mass associated with a specific number density at any redshift. 
If we assume that progenitor/descendant galaxy populations can be matched between different epochs at a constant comoving number density, we can infer the mass of Milky Way progenitor galaxies at higher redshifts by considering where the horizontal grey strip overlaps with the CMF at those redshifts as shown in Figure~\ref{fig:cum_number_density}.
This is the method commonly adopted when working with multi-epoch extragalactic observational data.
Here, this is achieved by inverting Equation~(\ref{eqn:CumMassFunc}) numerically using a Newton-Raphson root finding algorithm to solve for 
$ M_* =  M_*( z, N)$
given some constant choice of the galaxies' cumulative number density, $N$ as a function of redshift $z$.

\begin{figure*}
\centerline{\vbox{\hbox{
\includegraphics[width=3.5in]{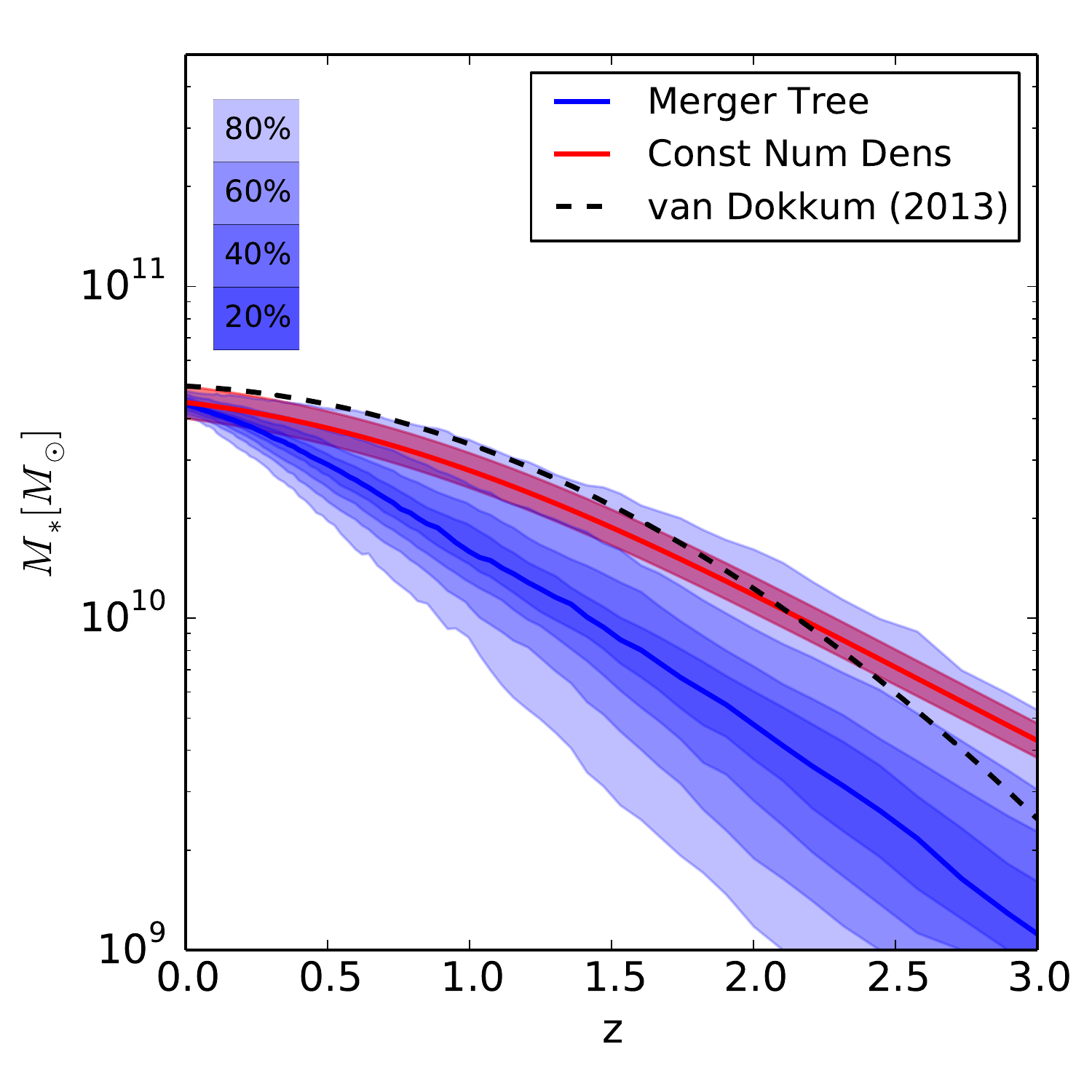}
\includegraphics[width=3.5in]{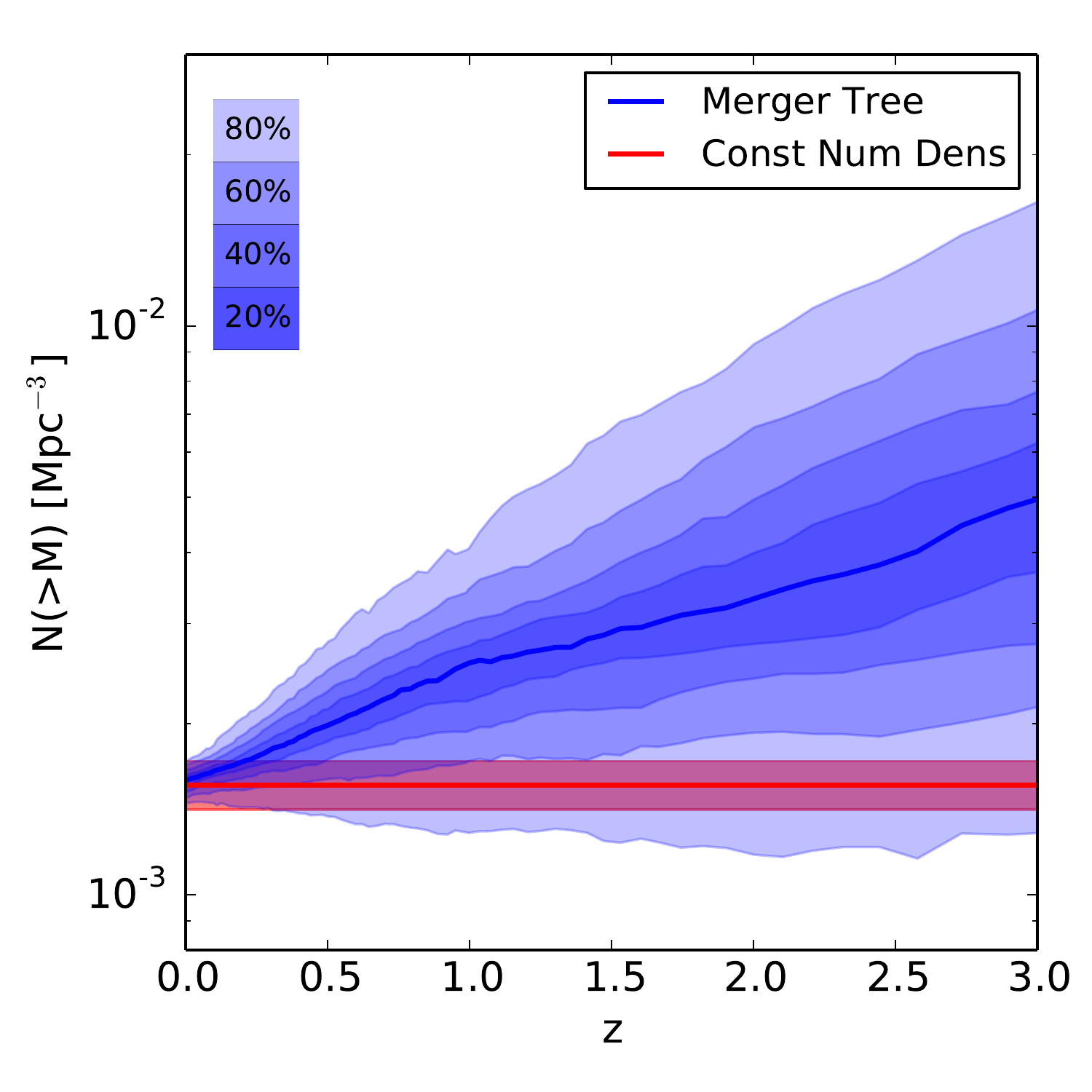}
}}}
\caption{(Left) The mass evolution of Illustris Milky Way mass galaxies is shown as a function of redshift for two methods used to trace galaxy mass growth with time.
The wide blue band indicates the mass distribution for a population of galaxies traced backward in time explicitly through the merger tree.
The red band indicates the inferred mass evolution by assuming constant comoving number density and applying the CMF fitting functions presented in Equation~(\ref{eqn:CumMassFunc}) with coefficients provided in Table~\ref{table:CMF_fit}.
The black dashed line indicates the observational inferred Milky Way mass evolution from~\citet{vanDokkum2013} using a constant comoving number-density assumption.
(Right) The number-density evolution of Milky Way mass galaxies is shown as a function of redshift.
The wide blue band indicates the distribution of number densities that a tracked galaxy population has when traced backward in time through the merger tree.
We note that there is both a median evolution with redshift and significant scatter, both of which are discussed in the text.  }
\label{fig:mw_mass_nd}
\end{figure*}

Second, the stellar mass evolution of the simulated galaxies can be measured directly from the galaxy merger tree, as was demonstrated in the previous subsection. 
The results from both mass tracking methods are presented in the left panel of Figure~\ref{fig:mw_mass_nd} with 
the red line indicating the median stellar mass evolution from number-density selection and
the blue solid line indicating the median stellar mass evolution from merger tree analysis.
The blue shaded regions show the range of stellar masses present from the tracked galaxy population, as noted in legend.
We note that, by definition, the blue line overlaps identically with the red line at the redshift where the galaxy populations is selected.

Two conclusions can be drawn from the left panel of Figure~\ref{fig:mw_mass_nd}:  
(1) The median mass evolution tracks based on the comoving number-density selection and direct merger tree tracking are offset from one another and 
(2) The merger tree tracking method possess significant scatter that grows as a function of redshift, indicating that there is a large diversity in the way galaxies assemble their mass, even when these are selected from a relatively narrow mass bin at $z=0$.

The stellar mass evolution for the explicitly tracked populations is more rapid than the inferred mass evolution from number-density selections.
By redshift $z=2(3)$ there is a factor of $\sim2(4)$ stellar mass offset between the two mass tracks.
The small scatter in the number-density selected mass evolution directly results from the assumption that galaxies remain rank ordered in their mass at all times.
In contrast, although all galaxies in the explicitly tracked galaxy populations are initially clustered in their stellar masses, the range of stellar mass values disperses with time as galaxies experience variable growth rates and stochastic evolution.

Also shown in the left panel of Figure~\ref{fig:mw_mass_nd} is a black dashed line corresponding to the Milky Way mass evolution derived from 3D-HST and CANDELS survey data using the constant comoving number-density selection in~\citet[][see their equation 1]{vanDokkum2013}.
A similar result was presented in~\citet{Patel2013}.
We find that -- though there are some differences in their detailed shape -- the simulated and observational comoving number-density mass evolutions are never separated by more than $\sim20\%$ over the redshift range $0<z<3$.
This is an indication that our simulated cumulative stellar mass function evolves quite similarly to the observed cumulative stellar mass function.
Any offset between the~\citet{vanDokkum2013} line and the simulation red curve are driven by inconsistencies between the two sets of mass functions.
However, more importantly, both of these mass evolution trajectories are offset from the explicitly tracked Milky Way mass evolution -- by a factor of $\sim2$ by redshift $z=2$.  
A factor of $\sim2$ median offset is not very severe~\citep[in agreement with the conclusions of][]{Leja2013} and is comparable to other uncertainties in stellar mass measurements (e.g., initial mass function uncertainties, age-dust degeneracies, weakly constrained star formation histories, etc.).
At the low mass end of the $z=2$ progenitor distribution, however, we find that roughly one third of explicitly tracked systems have masses that are offset by more than an order of magnitude from the constant comoving number-density mass trajectory.

The origin of the offset between the mass evolution tracks is simple:  galaxies do not remain exactly rank ordered, nor is galaxy number (density) a conserved quantity.
The right panel of Figure~\ref{fig:mw_mass_nd} shows the distribution of cumulative number density for the initially selected Milky Way type galaxies as they are traced back in redshift.
The dispersion of the galaxies' cumulative number densities with redshift is indicated through the dark blue bands, as noted in the legend.
The mass and number-density evolution shown in the left and right panels of Figure~\ref{fig:mw_mass_nd} are exactly interchangeable as long as we are using the cumulative stellar mass function to assign rank order. 
Thus, we reach the same two conclusions from the right panel as we did from the left: that the median number density of a tracked galaxy population evolves with time, and that the initially clustered population of galaxies disperses with time such that no single comoving number-density selection can fully recover the initial galaxy sample.
Correcting the median offset is fairly straightforward, and we provide a clear procedure for how to do so in the next subsection.

\begin{figure*}
\centerline{\vbox{\hbox{
\includegraphics[width=3.5in]{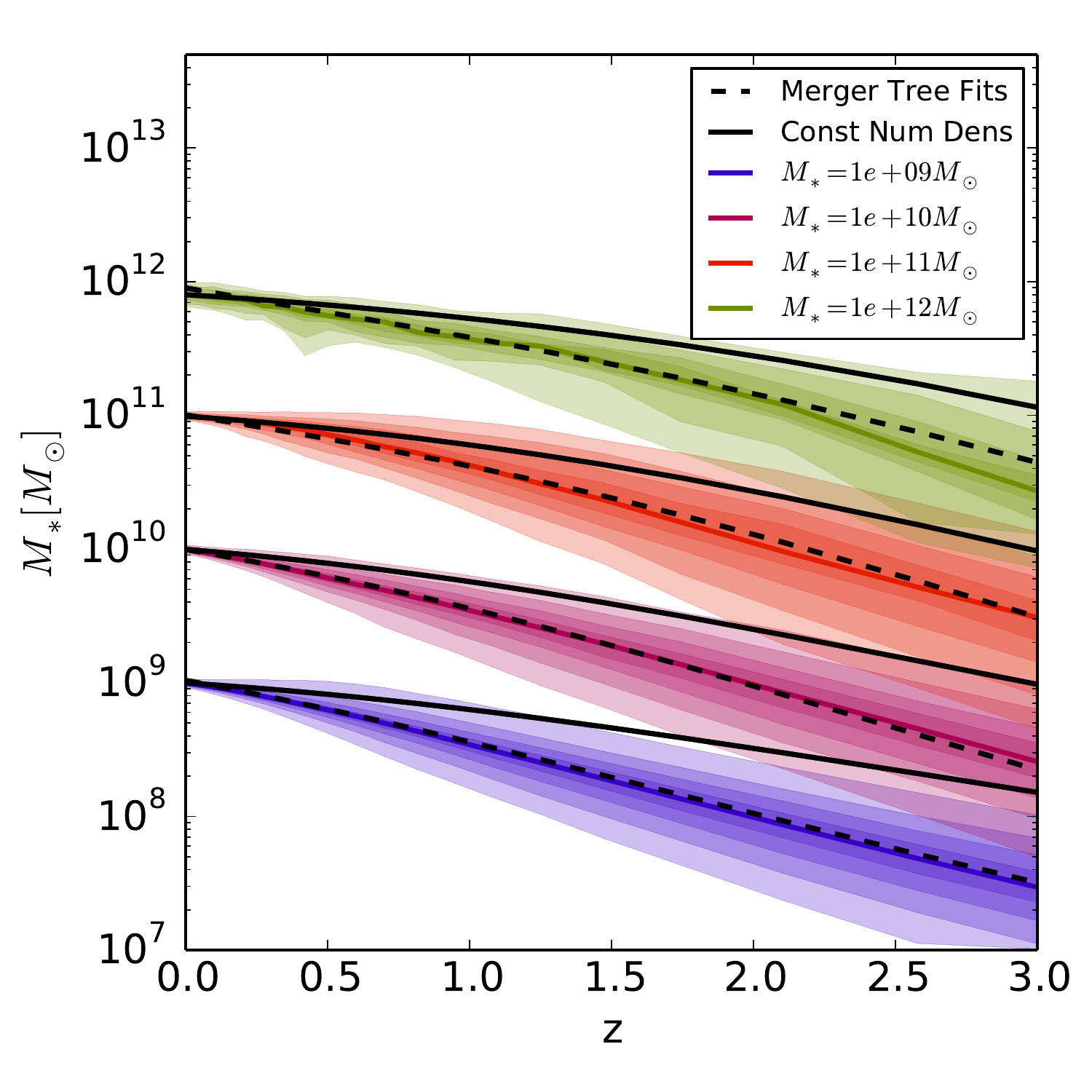}
\includegraphics[width=3.5in]{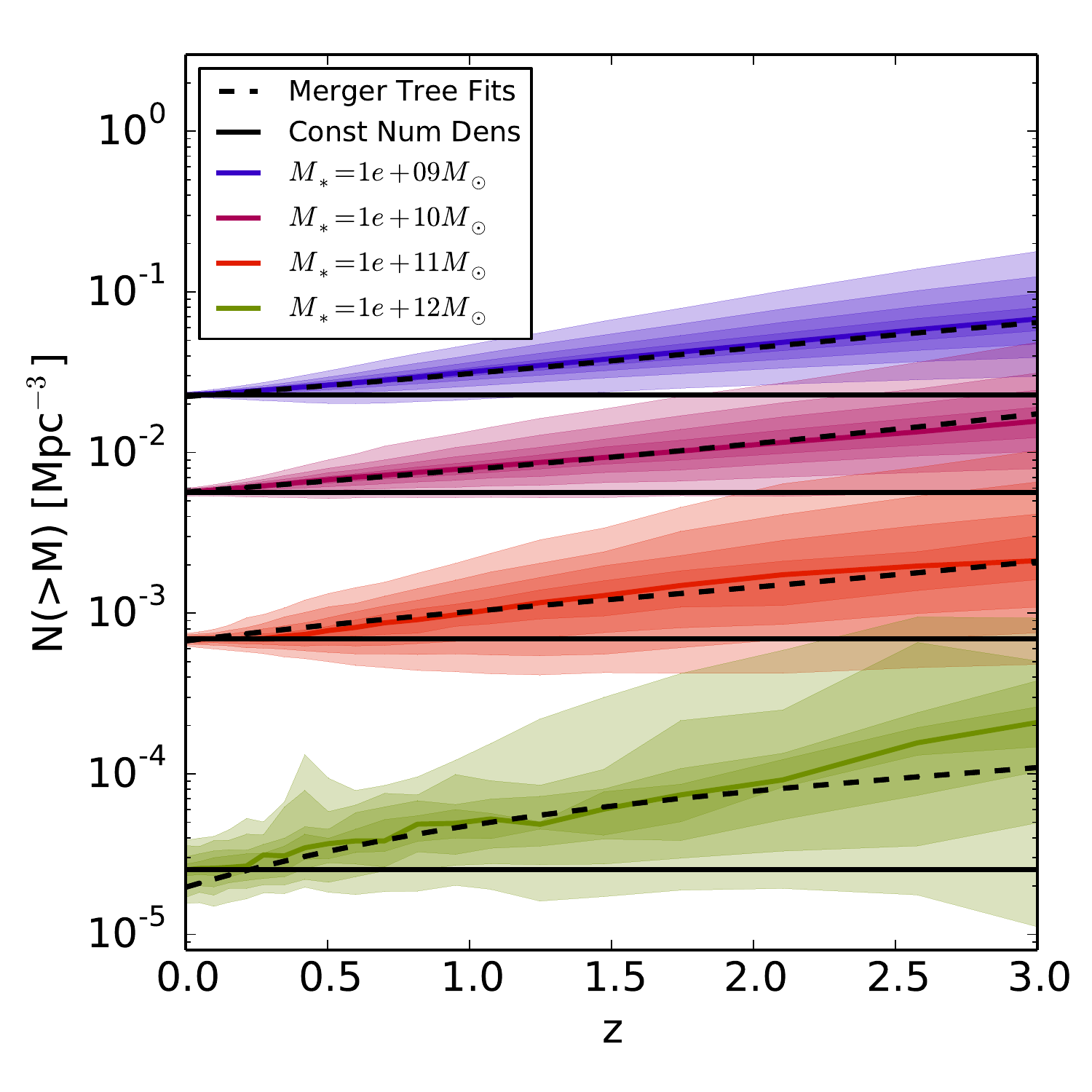}
}}}
\caption{
(Left) The mass evolution of $z=0$ galaxies tracked backward in time is shown as a function of redshift for three methods used to trace the stellar mass growth.
As in Figure~\ref{fig:mw_mass_nd}, the wide colored bands indicate the (20, 40, 60, 80)\% distribution of galaxies as tracked backward in time.
(Right) The corresponding number-density evolution is shown for the backward-tracked galaxy populations.
There is significant scatter in the number-density evolution which becomes worse for the most rare (lowest number-density) bins.
In both panels, black dashed lines indicate the results of the fit to the median evolution in number density, whose parameters appear in Table~\ref{table:Zevo_fit}.  
In the left panel, the inferred mass evolution is obtained by using the CMF to convert the best-fit number density to stellar mass.
Similarly, in both panels the solid black lines indicate the mass and number density evolution tracks following a constant comoving number density assumption.
}
\label{fig:multi_mass_nd}
\end{figure*}

\begin{figure*}
\centerline{\vbox{\hbox{
\includegraphics[width=3.5in]{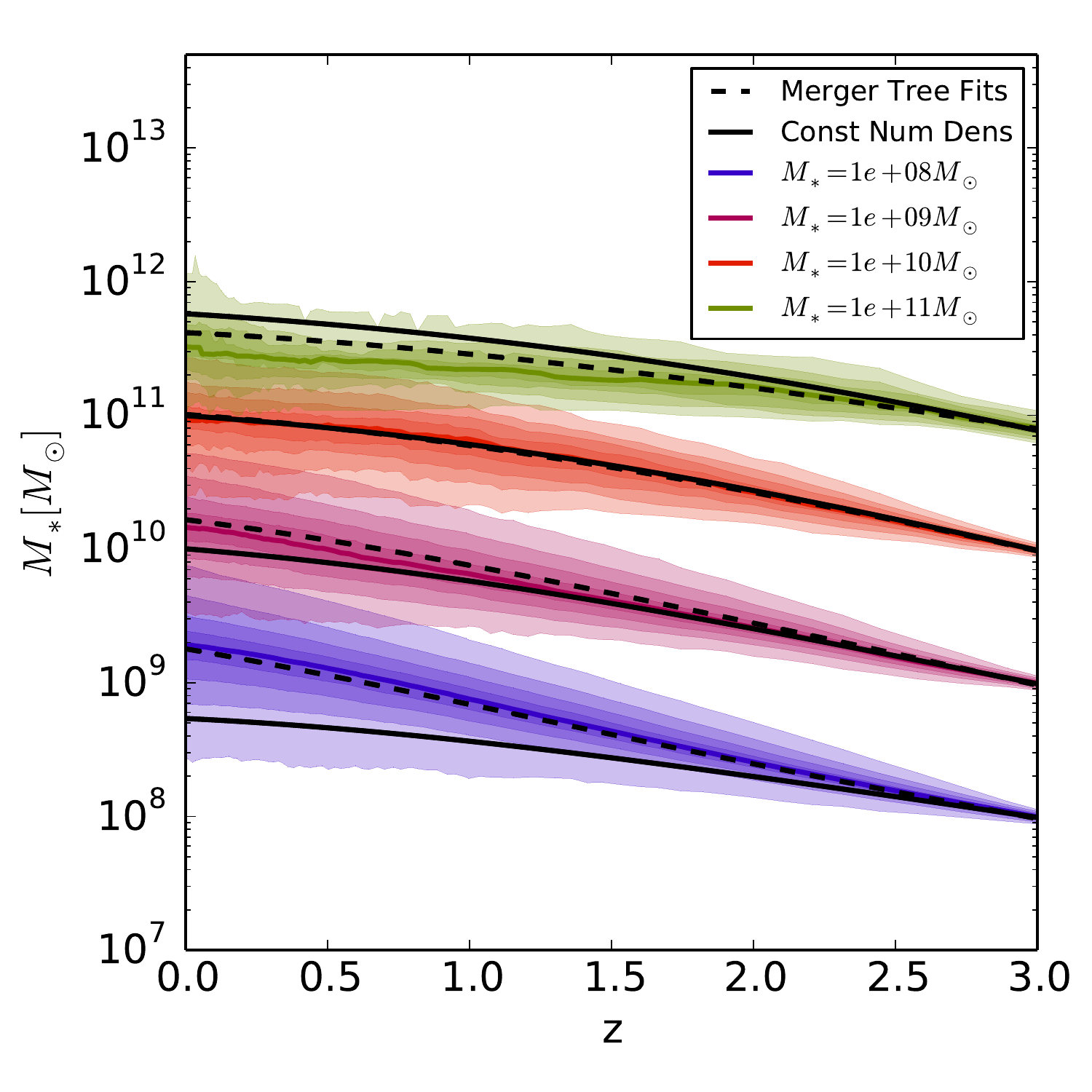}
\includegraphics[width=3.5in]{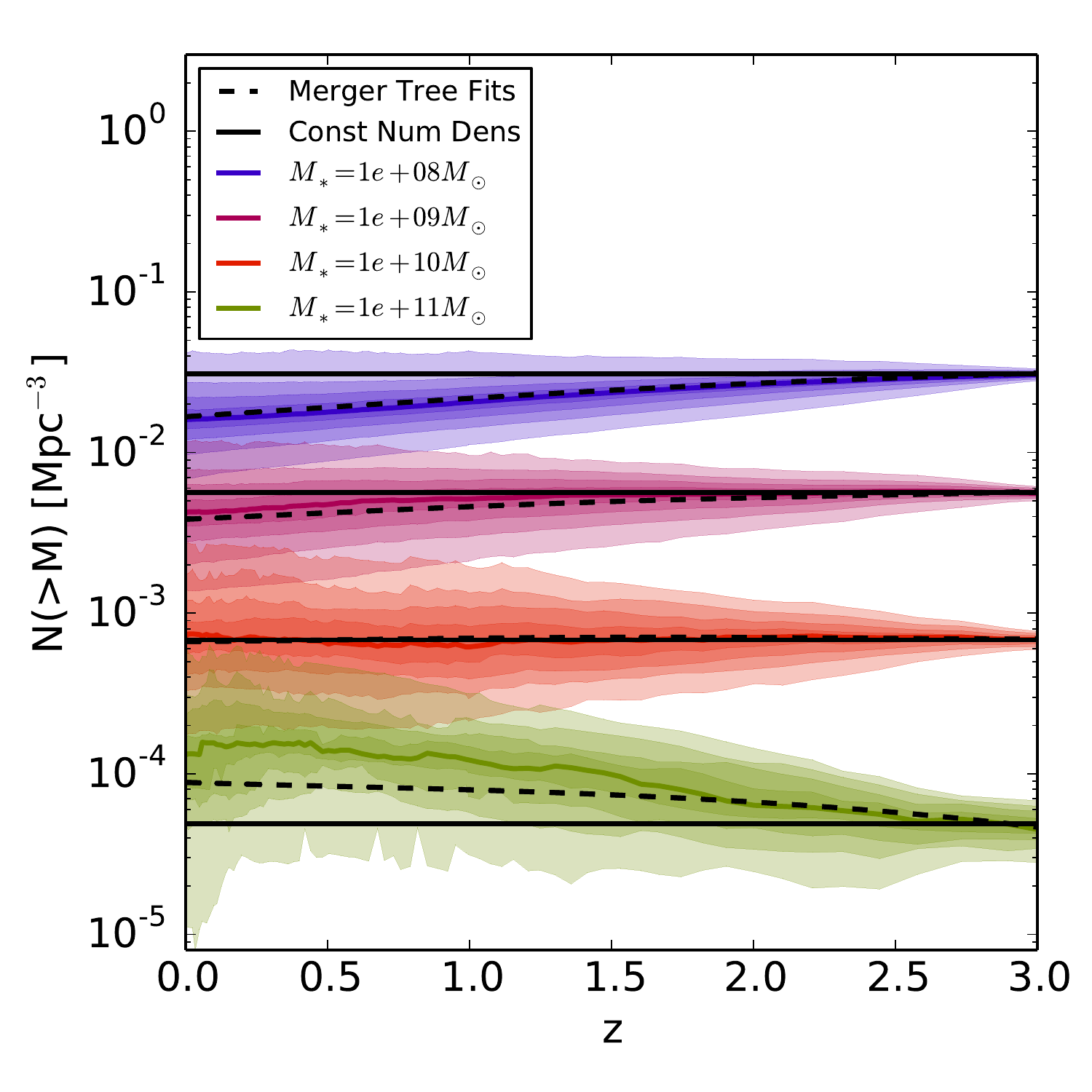}
}}}
\caption{
Same as Figure~\ref{fig:multi_mass_nd}, but with galaxy populations that are tracked forward in time after an initial selection at redshift $z=3$.  (Fit parameters appear in Table~\ref{table:fwd3}.)
Contrasting this figure with Figure~\ref{fig:multi_mass_nd} demonstrates the difference between tracking galaxy progenitor and galaxy descendant populations.
Whereas there is significant median evolution in the number density of backward tracked galaxy populations, we find that the overall number density evolution is somewhat weaker for forward tracked galaxy populations.  
This results in the constant comoving number density inferred mass evolution (solid black lines) more closely approximating the median merger tree tracked mass evolution (solid colored lines).
}
\label{fig:multi_rev_mass_nd}
\end{figure*}

\subsection{ Fits to the non-constant number density evolution across galaxy masses}
\subsubsection{Tracing Galaxies Backward in Time}
As described above, Equation~(\ref{eqn:CumMassFunc}) provides a redshift-dependent fit to the number density as a function of mass and redshift.
When we performed the regression to determine the best fit parameters in Section~\ref{sec:CMF}, we specified a set of $N$, $M_{*}$, and $z$ points based on data from cumulative stellar mass functions (i.e., the $N$ and $M_{*}$ pairings as shown in Figure~\ref{fig:cum_number_density}) at several redshifts.
This is sensible because this would be the only information that an observer would have on multi-epoch galaxy populations.
However, since the cumulative stellar mass functions are built {\it independently} at each redshift, no information regarding the number-density evolution of individual galaxies or populations of galaxies is retained using this approach.
As a result we find that when we track a population of galaxies explicitly in time they follow a non-constant number-density evolution in time as shown in Figure~\ref{fig:mw_mass_nd}.

Using information on the mass and number-density evolution as a function of redshift we can perform a similar regression analysis using Equation~(\ref{eqn:CumMassFunc}).
However, here we want to find the cumulative number-density evolution track that best describes the actual tracks taken by galaxies in our simulation.
To achieve this we take $N=N(z)$ to be the number density of {\it each individual galaxy} as it evolves in time from an initial $z=0$ mass of $M_{*,0}$.
The only difference between the fit that we perform in this section and what we did in the previous subsection is the way that the ($N$,$M_{*}$,$z$) pairings are constructed.
Here, rather than using the pairings from the CMF, which are constructed independently at each redshift, 
we use ($N$,$M_{*,0}$,$z$) pairings constructed from the merger tree by tracing each galaxy with redshift $z=0$ mass, $M_{*,0}$, backwards in time to obtain the number density evolution for every galaxy $N=N(z)$.

The derived parameters from this fitting procedure are given in Table~\ref{table:Zevo_fit}.
Using this fit we are able to infer the expected median number density a galaxy population will have at some redshift $z$ given its initial $z=0$ mass, $M_{*,0}$.
Using this fit in conjunction with the tabulated CMF presented in Section~\ref{sec:CMF} we can then infer the average mass associated with this galaxy population at other redshifts.

\begin{figure}
\centerline{\vbox{\hbox{
\includegraphics[width=3.4in]{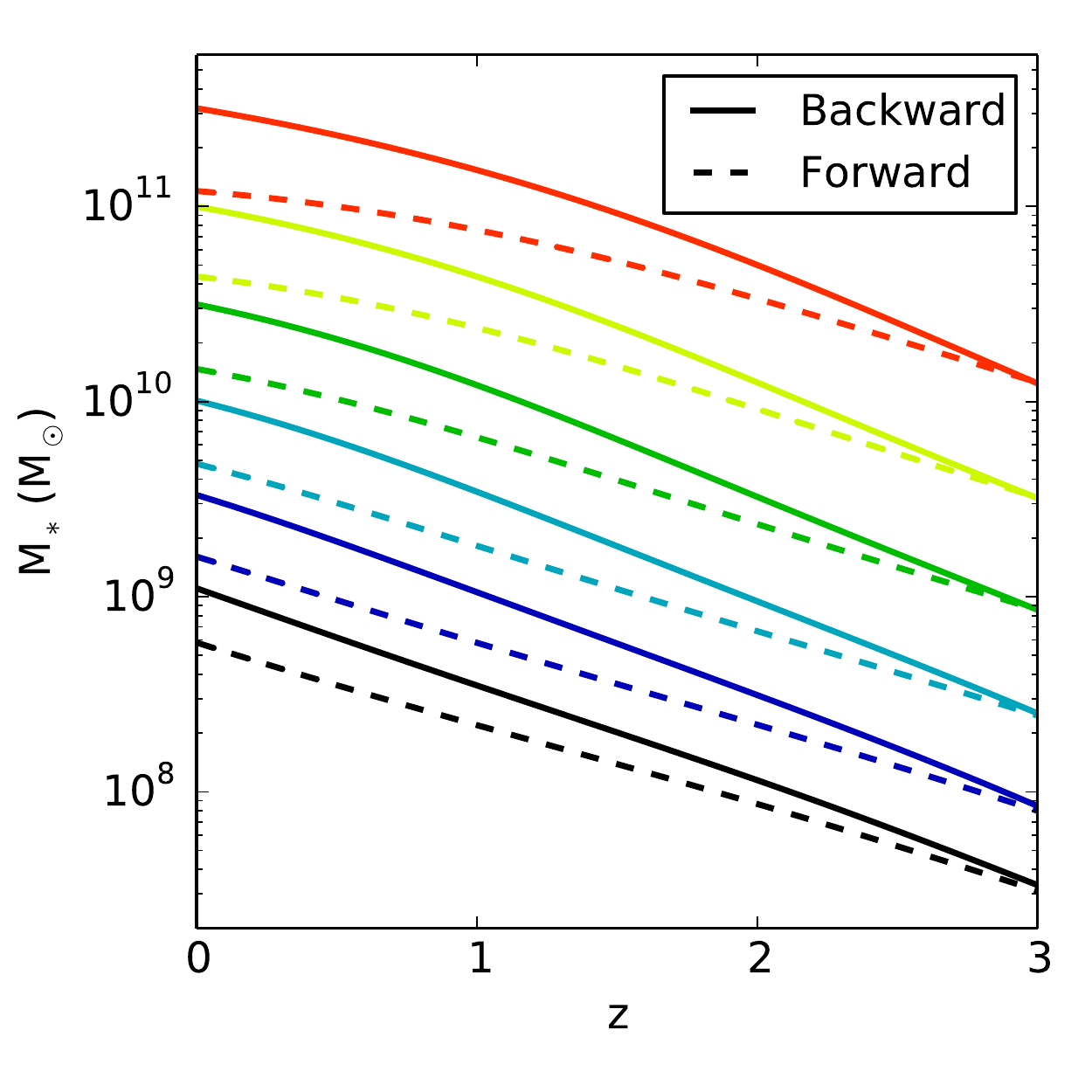} }}}
\caption{ The average mass evolution is shown for several mass bins to contrast the results that are obtained from tracking galaxies forward and backward in time.  
Both galaxy merger events and asymmetric median galaxy growth rates cause the inferred mass evolution to be different at the factor of $\sim$2 level depending on the directionality of the galaxy tracking.
Note that neither of these curves is fundamentally more correct than the other, but rather they identify different mass evolution tracks as described in more detail in the in Sections~\ref{ref:fow_back_comp} and~\ref{sec:tracking_asymmetry}. }
\label{fig:forward_backward_comp}
\end{figure}

We demonstrate the derived mass and number-density evolution in the left and right panels of Figure~\ref{fig:multi_mass_nd}, respectively.
Figure~\ref{fig:multi_mass_nd} shows the mass and number-density evolution of a set of four different galaxy populations where the colored bands are constructed from tracking galaxies along their main progenitor branch, as described above.
All of the same behavior that was present for the Milky Way mass bin inspected above is also seen for the other galaxy mass bins considered here.
The black solid lines in each panel indicate the mass and number density evolution that is obtained from the constant comoving number density assumption.
The black dashed lines in each panel indicate the mass and number-density evolution that is obtained from the non-constant comoving number density fit.

By construction, we find that the non-constant number-density fit follows the appropriate average trend.
We obtain the mass evolution shown in the black dashed line left panel of Figure~\ref{fig:multi_mass_nd} by converting the evolving number density into a mass based on the tabulated CMF coefficients given in Table~\ref{table:CMF_fit}.
This procedure can be replicated with observational data.
The fit parameters from Table~\ref{table:Zevo_fit} are well suited to describe the median number-density evolution over the resolved mass range $ M_*(z=0) > 10^9 M_\odot$ and redshift range $0< z < 3$.
The fit parameters given in Table~\ref{table:Zevo_fit} can be applied to identify the progenitor galaxies that properly follow the median mass evolution of an initially selected galaxy population~\citep[e.g.,][]{Behroozi2013, Marchesini2014}.
However, we emphasize that while this fit does describe the median number-density evolution, it does not capture the scattered growth rates.

\subsubsection{Tracing Galaxies Forward in Time}
\label{ref:fow_back_comp}

An important caveat for the fit parameters presented in Table~\ref{table:Zevo_fit} is that they were obtained by identifying galaxies at $z=0$ and tracing their mass/number density backwards in time, and therefore only apply in that direction.
This informs us of the mass and number-density evolution of the main progenitor galaxies of the galaxy population that is present at redshift $z=0$.
However, this analysis is not an inherently reversible procedure owning primarily to asymmetric scattered growth rates when examining the forward and backward evolutionary paths of galaxy populations. 
Therefore, while the results of the previous section can be used to identify the past mass or number-density evolution of present day selected galaxy populations, 
we cannot necessarily use the results of the previous subsection to identify the present day counterparts to an observed high-redshift galaxy population.
Instead, to identify the present-day descendants of a high-redshift galaxy population, the analogous galaxy populations need to be traced forward in time.

To perform this analysis, we select a galaxy population at some non-zero redshift and use the merger trees to follow that galaxy population forward in time.
Galaxies that merge with more massive systems are followed until the merger event, after which we assume that the galaxy is no longer observable -- and so it is not included in the fitting or analysis beyond that point.
Neglecting these branches entirely does not significantly change the results, but the consumption of galaxies via mergers can lead to a significant reduction in the number of galaxies that are available for forward tracking.  
We find a similar result for the ``survival fraction'' of galaxies to $z=0$ as has been shown in previous work~\citep[e.g.,][]{Leja2013, Mundy2015}.
Specifically, the survival fraction is not a very steep function of initial stellar mass, with only a weak trend where more massive systems are more likely to survive.
For a galaxy population selected at redshift $z=2$, roughly two-thirds of those galaxies can be expected to have redshift $z=0$ counterparts with the rest having been consumed by some larger system.  
For a $z=3$ selected galaxy population, the survival fraction drops to roughly half.
This can be contrasted with the expectation that 100\% of redshift $z=0$ selected galaxies have meaningful high-redshift main progenitors, which only requires assuming that the employed simulation has sufficient resolution to continue to track their formation backward in time.

Figure~\ref{fig:multi_rev_mass_nd} shows the mass and number-density evolution of a population of galaxies selected at redshift $z=3$ and tracked forward in time.
The mass evolution shown in Figure~\ref{fig:multi_rev_mass_nd} is qualitatively consistent with our basic expectations:
the median galaxy mass increases with time along with an increase in the dispersion of the individual galaxy mass distribution.
We perform a regression on the ($N$, $M_{*,z}$, $z$) pairings to determine the coefficients of Equation~(\ref{eqn:CumMassFunc}), 
where we take $M_{*,z}$ to be the stellar mass of each galaxy at some initial redshift $z$ (in place of $M_{*,0}$ from the previous subsection), and $N=N(z)$ is the mass ranked cumulative number density of each galaxy when traced forward in time.
We indicate the fit mass and number-density evolution tracks with black dashed lines in both panels of Figure~\ref{fig:multi_rev_mass_nd} and the best fit coefficients are given in Tables~\ref{table:fwd1}-\ref{table:fwd3}.
We find that the lower three mass bins are tracked very well in time using this fit.
The highest mass bin shows some significant deviation from the fit which is mostly a consequence of the low number of galaxies in this bin -- which decreases as it moves to lower redshift.

Interestingly, we find that the qualitative behaviors of the number-density evolution for the tracked galaxy populations are different when tracked forward and backward in time.
When tracked backward in time (Figure~\ref{fig:multi_mass_nd}), the number density of the tracked galaxy population steadily increases.  
However, when tracked forward in time the median number density of the tracked galaxy population remains much more constant.
For comparison, the solid black lines in Figure~\ref{fig:multi_rev_mass_nd} indicate evolutionary tracks of constant comoving number density.
We find that the median mass and number density evolution for one of the bins -- the red band, which was selected to contain galaxies with stellar masses of $M_*=10^{10} M_\odot$ at redshift $z=3$ -- almost identically follows the constant comoving number density trajectory.
The other tracked bins are offset from the constant comoving number density track, but remain closer to this constant comoving number density track than their backward tracking counterparts presented in Figure~\ref{fig:multi_mass_nd}.

If we compare the resulting mass evolution using the backward and forward fits, as shown in Figure~\ref{fig:forward_backward_comp},
we find that tracing galaxies forward in time yields a noticeably shallower mass evolution.
The differences in the forward/backward number-density evolution as well as the offsets in the forward/backward inferred mass growth rates are both 
primarily driven by the scattered growth rates of galaxies, as we discuss in Section~\ref{sec:tracking_asymmetry}.

\section{Results:  Tracing Galaxies via Stellar Velocity Dispersion}
\label{sec:Results_VDF}
\subsection{Cumulative Velocity Dispersion Function}
Velocity dispersion has been advocated as a stable proxy for galaxy rank order because of its invariance to growth via galaxy mergers~\citep[e.g.,][]{Loeb2003, Bezanson2011}.
If galaxy growth is driven primarily by mergers then the central velocity dispersion will evolve by $\leq30\%$ by redshift $z=3$~\citep{Hernquist1993, Hopkins2009b}.
In contrast, if internal changes~\citep[e.g., puffing up via mass loss from quasars;][]{Fan2008} dominate over mergers in determining structure evolution of low redshift galaxy populations, then the velocity dispersion can increase significantly.
While this point has been examined through semi-analytic models in the past~\citep{Leja2013, Mundy2015}, it has not previously been inspected using numerical simulations where the velocity dispersion of galaxies can be tracked directly.
In parallel with the previous section, here we present the multi-epoch cumulative velocity dispersion function (CVDF) along with a multi-epoch simple fitting function to determine its ability to reliably link galaxy populations in time.

\begin{figure}
\centerline{\vbox{\hbox{
\includegraphics[width=3.5in]{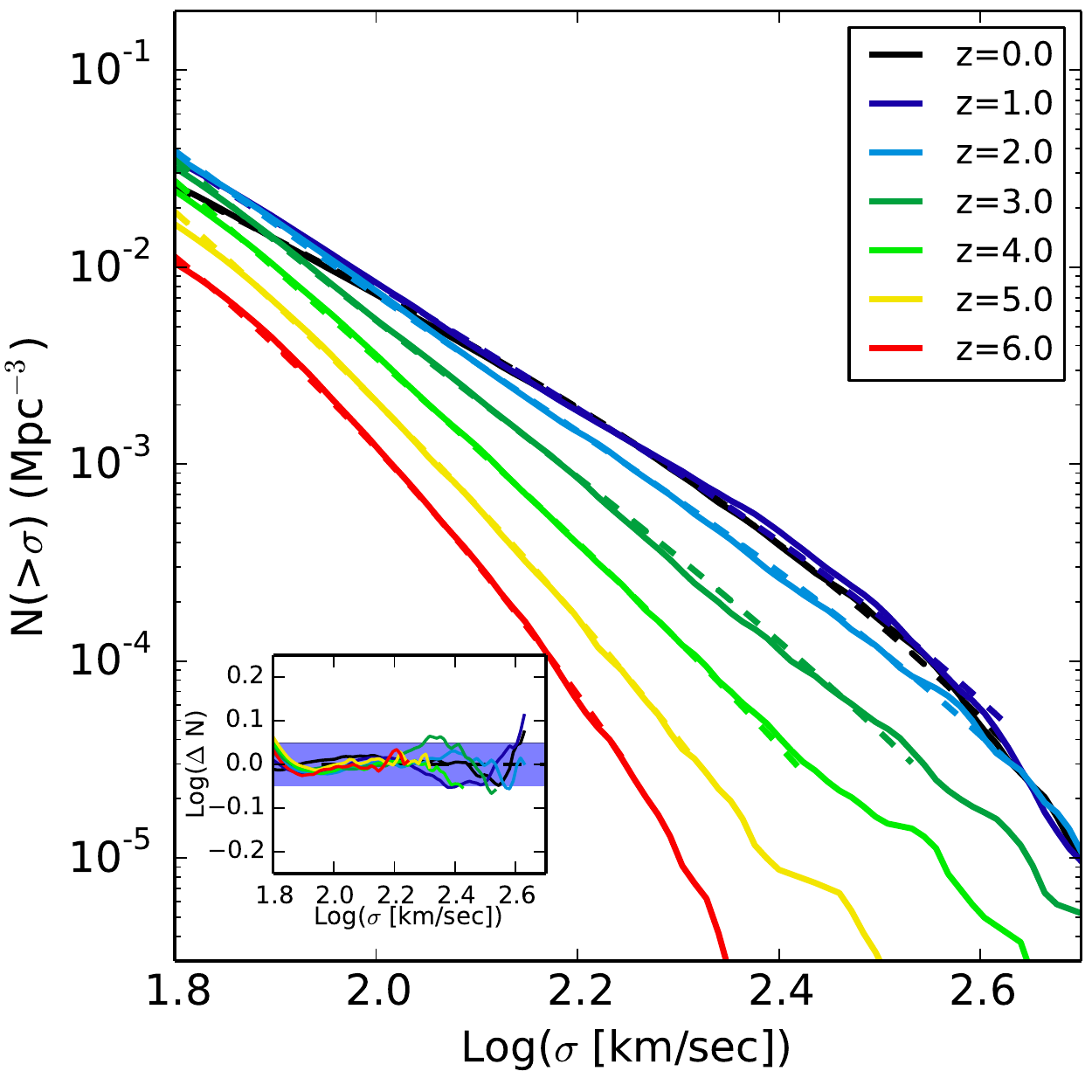}}}}
\caption{ The cumulative velocity dispersion function (CVDF) is shown for several redshifts, as indicated in the legend.
In contrast to the CMF, the CVDF shows comparatively little evolution with redshift after $z=2$.
Multi-epoch fits given in Equation~(\ref{eqn:CMVDFunc}) are indicated with dashed lines, with the errors associated with these fits indicated in the inset plot.
Fit parameters can be found in Table~\ref{table:CVDF_fit}.}
\label{fig:CVDF}
\end{figure}

To construct the cumulative velocity dispersion function, we define the central stellar velocity dispersion, $\sigma_*$, as the three-dimensional standard deviation of the stellar velocities within the stellar half-mass radius.
We present the CVDF at several redshifts in Figure~\ref{fig:CVDF}.
We employ a fit to the CVDF of the form
\begin{equation}
N(\sigma_*) =  A \; \tilde{ \sigma}_* ^{\alpha + \beta \mathrm{Log}\tilde{ \sigma}_*  } \; \mathrm{exp}({-\tilde{ \sigma }_* }) 
\label{eqn:CMVDFunc}
\end{equation}
where $\tilde{ \sigma}_* = \sigma_* / 10^\gamma {\rm km/sec} $.
We perform a regression analysis to determine the above coefficients ($A$, $\alpha$, $\beta$, and $\gamma$), each of which contains a second order redshift dependence as given in Equations~\ref{eqn:A}-\ref{eqn:E}.
The derived coefficients are given in Table~\ref{table:CVDF_fit} with the results shown as dashed lines in Figure~\ref{fig:CVDF}.
The inset axes show the error associated with the multi-epoch fit, which is accurate at the $\ltsim10$ percent level between redshifts $0<z<6$ over the velocity dispersion range ${\rm Log}(\sigma_* {\rm km/sec} ) > 1.8$ and number density range $N>3\times10^{-3} {\rm Mpc^{-3}}$.

We find that there is relatively limited evolution in the CVDF from $z=2$ to $z=0$. 
This low level of redshift evolution was not present for the CMF for any mass scale.
We do find that there is evolution in the CVDF beyond redshift $z=2$ which can be quite significant at all velocity dispersion values.

 \begin{figure*}
\centerline{\vbox{\hbox{
\includegraphics[width=3.5in]{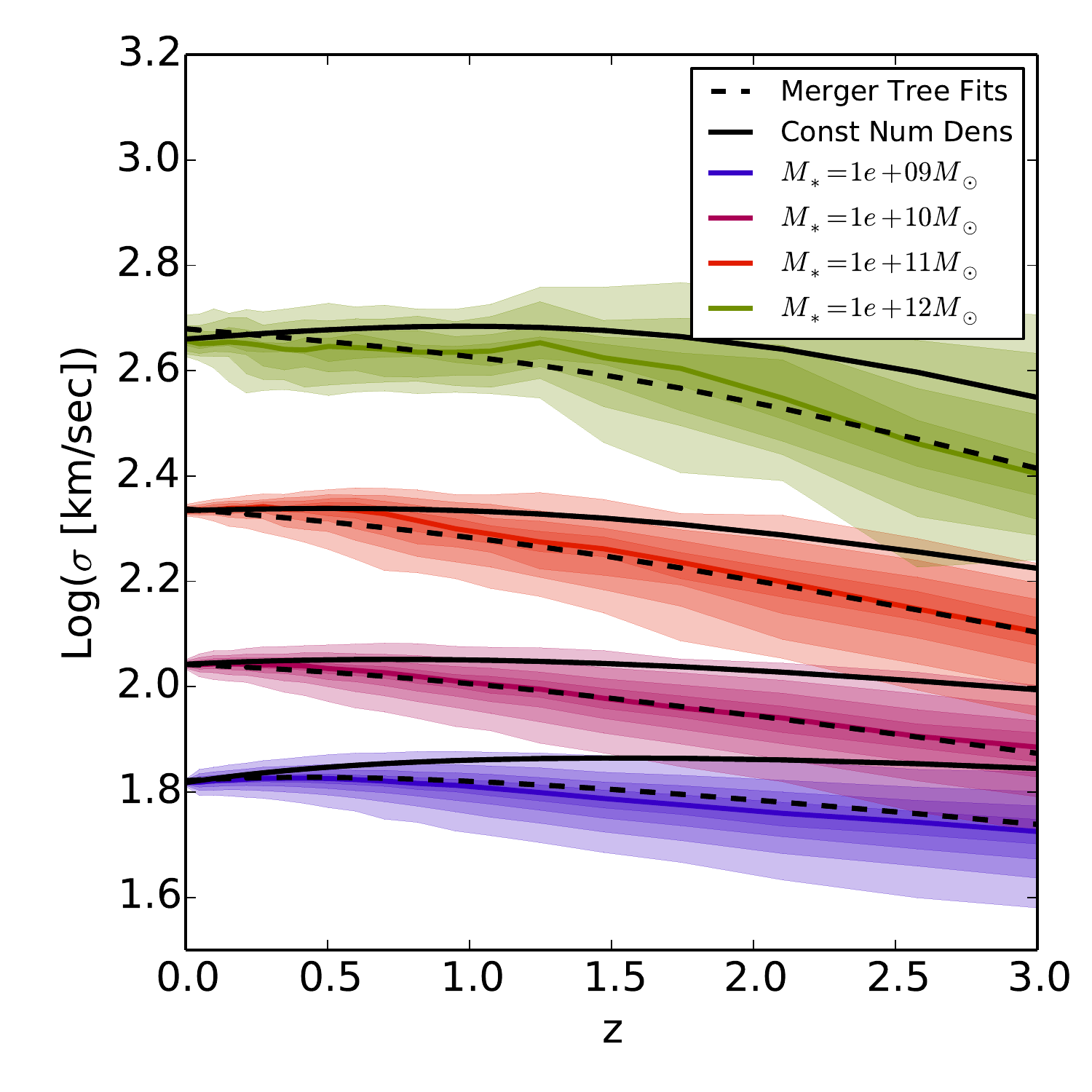}
\includegraphics[width=3.5in]{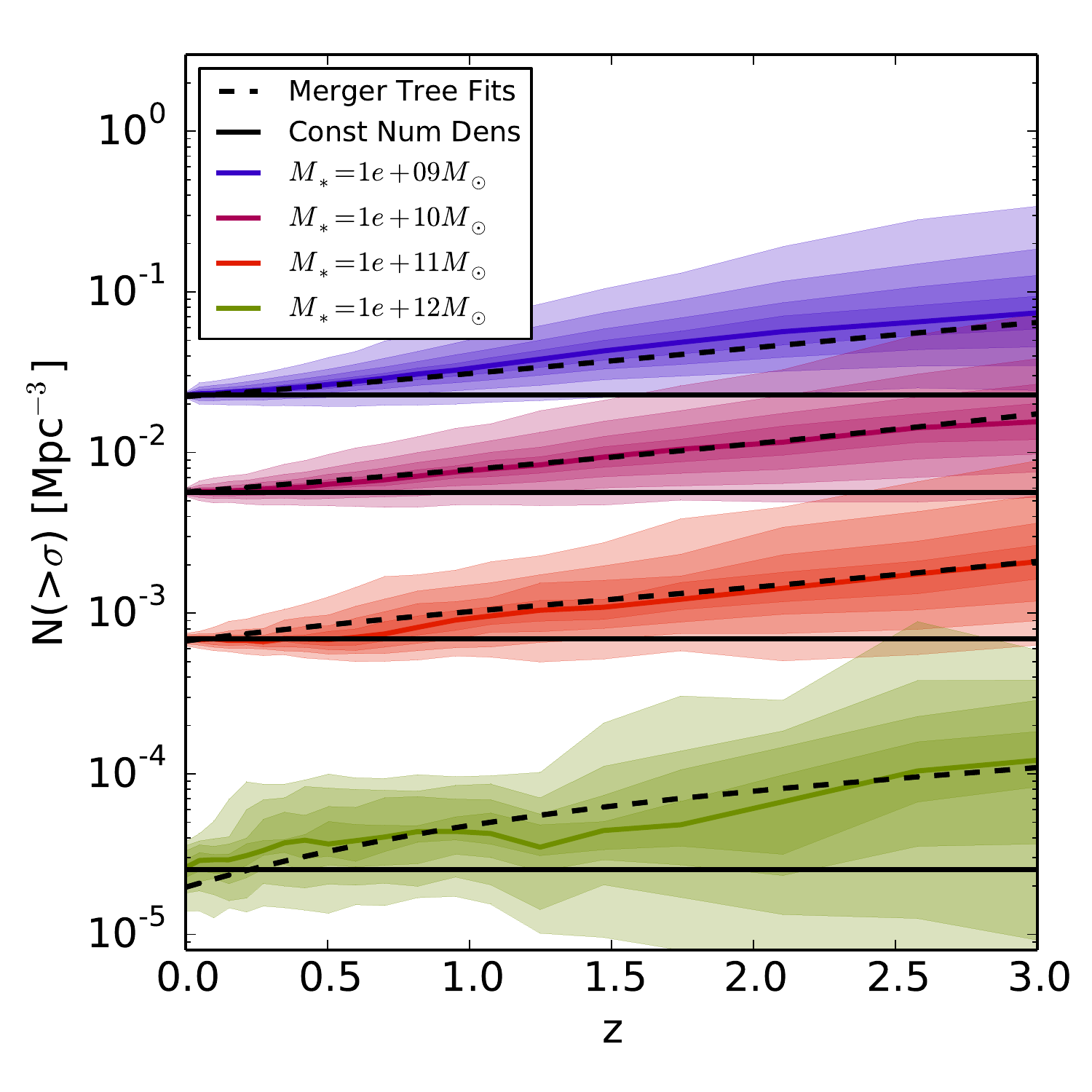}
}}}
\caption{
Analogous to Figure~\ref{fig:multi_mass_nd}, but where we are now selecting galaxies in bins of number density according to stellar velocity dispersion rather than stellar mass.
The initial number density ranges sampled in each bin are chosen to match what was used in Figure~\ref{fig:multi_mass_nd}.
The legend indicates the stellar mass that corresponds to each number density bin.
We find that galaxies do not show very significant median evolution in their velocity dispersion when traced out to redshift $z=3$.
However, we find there is significant spread associated both in terms of the velocity dispersion and number-density distribution for this galaxy population when traced backward in time.
The black dashed lines correspond to the same number-density evolution fit shown in Figure~\ref{fig:multi_mass_nd} which was constructed using the CMF, {\it not} a new fit using the CVDF.  
In the left panel, the CVDF was used to convert number density to velocity dispersion.  
Despite having been constructed using stellar mass, the evolving number density fit also appropriately follows the velocity dispersion evolution.
   }
\label{fig:multi_vd_evo}
\end{figure*}

\subsection{Evolutionary Tracks in Velocity Dispersion}
The limited evolution in the CVDF is an intriguing feature for comoving number-density analysis.  
If the CVDF is assembled at early times, then perhaps the central velocity dispersion evolution is restricted in time.
This would happen in a scenario where galaxies attain their central velocity dispersion at early times without significant evolution thereafter, even in the presence of mass growth.
This is expected for massive quenched galaxies that assemble at early times and retain their internal stellar structure while ``puffing up" at late times from minor merger events~\citep{Naab2007}.
This scenario is less likely to apply to star forming galaxies.

We test the velocity dispersion rank ordering by tracing several galaxy populations back in time, similar to what was done in the previous section.
The results are shown in Figure~\ref{fig:multi_vd_evo}.
We could select the exact same systems used in the previous section to trace backwards in time.
However, because there is not a perfect 1:1 correlation between stellar mass and velocity dispersion, using the same mass-selected galaxy population would introduce a somewhat larger initial spread in the velocity dispersion-assigned cumulative number density (the introduced extra scatter in number density is roughly a factor of 3).
Given our goal of understanding how cumulative number-density selection methods are able to trace galaxy populations in time, we instead select a galaxy population that uses the same initial number-density limits as in the previous section, but use the CVDF rather than the CMF to assign number density.
For the velocity dispersion bin centered around a number density of $6\times10^{-3}$ Mpc${}^{-3}$, this results in the selection of 466 galaxies, 130 of which were also in the mass-selected galaxy sample in the same number density range.
Although this is a somewhat different initial galaxy selection, we can consider the evolution of this galaxy population in velocity dispersion and directly compare the evolution in number-density space against what we found in the previous section.

The velocity dispersion evolution is shown in the left panel of Figure~\ref{fig:multi_vd_evo}.
The median velocity dispersion evolution is shown with the solid colored lines, while the shaded colored regions identify the spread in the evolving velocity dispersion distribution for the initially selected galaxy population as indicated in the legend.
We find that the median velocity dispersion does not significantly change over this period of time for any of the bins.
A typical change of 0.1-0.2 dex from redshift $z=0$ out to $z=3$ is found.
For comparison, the solid black line indicates the evolution along an assumed constant comoving number density trajectory.

However, we find that the mild evolution of the velocity dispersion {\it does not} directly translate to the proper recovery of the initial galaxy population when selected via their comoving number density.
The right panel of Figure~\ref{fig:multi_vd_evo} shows the number-density evolution as assigned from the CVDF for this tracked galaxy population.
Despite the mild evolution of the velocity dispersion, the divergence of this galaxy population in number-density space with time is still significant.
There is a median offset in the number density of this tracked galaxy population that grows with time and the scatter of the initial galaxy population reaches order of magnitude or larger levels in comoving number density by redshift $z=3$.
The general trend that we find in the evolving number-density distribution is very similar in terms of median offset and scatter growth to that found when we used the CMF to trace galaxies in time.
To highlight these similarities, the black dashed line in Figure~\ref{fig:multi_vd_evo} {\it is not a new fit from this data}, but rather the number-density evolution determined in the previous section using the CMF (i.e., the coefficients given in Table~\ref{table:Zevo_fit}).
We find that the median number-density evolution that we derived for the CMF applies very well to the CVDF number-density evolution.

Given the applicability of the CMF number-density evolution fit, we can consider the inferred velocity dispersion evolution.
Specifically, we can calculate $N=N(z)$ using the fits from the previous section and then determine $\sigma_*=\sigma_*(N(z))$ via Equation~(\ref{eqn:CMVDFunc}) with the coefficients given in Table~\ref{table:CVDF_fit}.
The result is shown as the black dashed line in the left panel of Figure~\ref{fig:multi_vd_evo}, which is in good agreement with the merger tree tracked velocity dispersion evolution.
This is a clear indication that there is a mean evolution in the number density of galaxies present at nearly the same level regardless of whether we employ the CMF or CVDF to Milky Way mass systems.

\section{Discussion}
\label{sec:Discussion}
The method of matching galaxies between different epochs observationally based on their number density is both widely used and reasonably physically justified.
As has been shown in the past -- and as we have confirmed in this paper -- the errors that are introduced into the inferred stellar mass evolution of galaxies between redshift $z=2$ and the present day when one uses a constant number-density selection rather than the explicit galaxy merger tree mass evolution are not catastrophic (i.e. of order $\sim0.3-0.5$ dex).
By neglecting the scattered growth histories of galaxies, one can immediately link high and low redshift observed galaxy populations.
This is one of the principal methods employed to infer galaxy mass buildup, size growth, and morphology evolution in past literature.
However, such an approach does not properly link the vast majority of progenitor and descendant galaxies~\citep{Mundy2015}.
Depending on the elapsed time and mass/number-density bin size, the true recovery rate of progenitor/descendant galaxies using a constant comoving number-density selection can easily be of order $\sim10-30\%$~\citep{Leja2013, Mundy2015}.

A crude link does exist between high- and low-redshift galaxies in their comoving number density, but this link evolves with time and includes significant intrinsic scatter.
In this paper we have presented the explicitly tracked number-density evolution of galaxies based on a hydrodynamical simulation of galaxy formation.
We find a median offset associated with the growth history of any galaxy population when compared against the constant comoving number-density selection methods.
The magnitude of this offset is not the same when tracking galaxies forward and backward in time.
We find that tracking galaxies forward in time yields median mass and number density evolution tracks that evolve in better agreement with the constant comoving number density than when systems are tracked backward in time.
We have provided simple fitting functions that describe the median number-density evolution -- both forward and backward in time -- that can be applied to observational studies straightforwardly.
Once we adopt a simple formulation for the non-constant comoving number-density evolution, we can recover the median mass evolution of our explicitly traced galaxy population from the CMF alone.
We encourage this to enter into future observational analysis as has been done in~\citet{Marchesini2014}.

We have examined the claim that velocity dispersion can act as a more robust property for linking galaxies together in time.
By constructing the CVDF, we were able to apply an identical analysis to the evolution of the velocity dispersion of our tracked galaxy population.
Although the CVDF itself shows limited evolution from $z=0$ to $z=2$, there is still significant evolution in the number density of individual galaxies as assigned through the CVDF.
We found that the median evolution in the number density for a population of explicitly-tracked galaxies behaved nearly identically to what we found when we used the CMF.
This illustrates an important point:  it indicates that there is an underlying driver of galaxy number-density evolution that impacts our results regardless of the physical quantity on which we perform our galaxy rank ordering.

\subsection{Dependence on Baryon Physics}
\begin{figure*}
\centerline{\vbox{\hbox{
\includegraphics[width=3.5in]{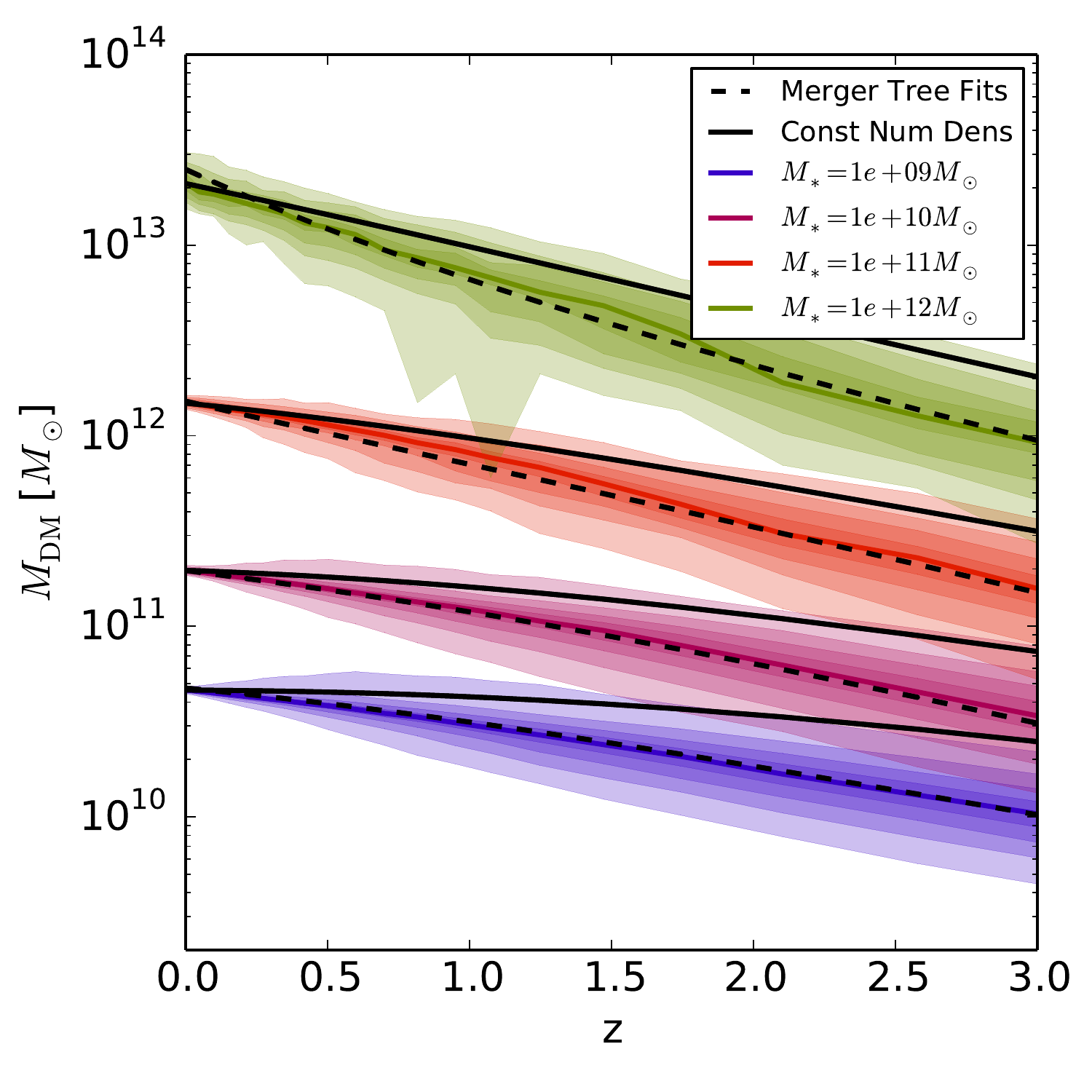}
\includegraphics[width=3.5in]{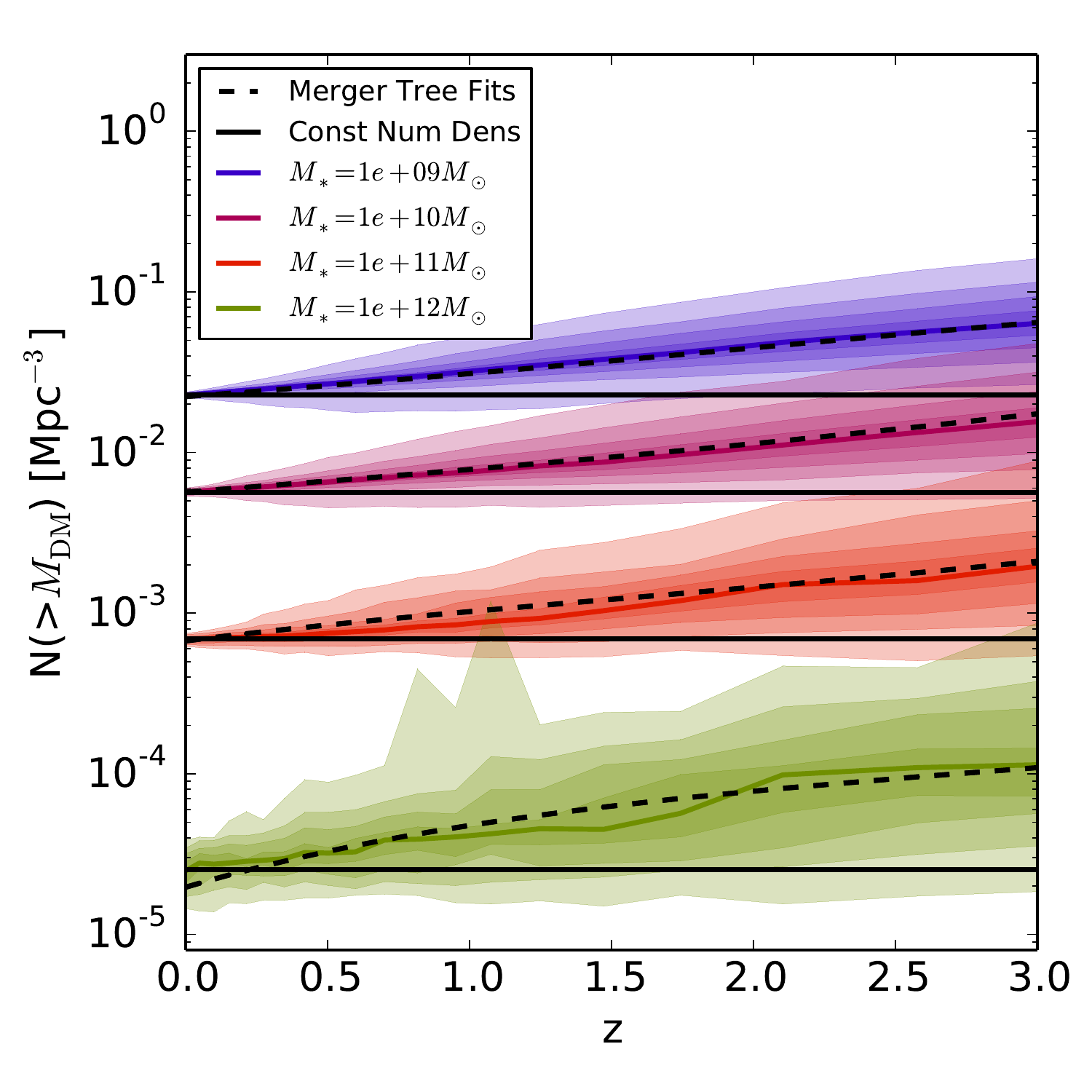}
}}}
\caption{ Analogous to Figures~\ref{fig:multi_mass_nd} and \ref{fig:multi_vd_evo}, but where we are now selecting galaxies in bins of number density according to their dark matter subhalo mass rather than stellar mass or velocity dispersion.
As in Figure~\ref{fig:multi_vd_evo}, the black dashed lines are number-density evolution tracks constructed using the CMF, and the cumulative dark matter mass function was used to convert number density to dark matter mass in the left panel.
We find that the dark matter mass growth of these systems looks somewhat different from the stellar mass growth owing to the lack of quenching.  
Regardless of this difference, the number density evolution is nearly identical to what we obtained for both the central velocity dispersion and stellar mass number density analysis, and the fit correctly tracks the dark matter mass evolution.}
\label{fig:multi_dm_evo}
\end{figure*}

How much do the prescriptions derived for the galaxy comoving number density evolution in this paper depend on the specific physics implementations that we have employed in the Illustris simulation?
The answer to this question is fairly critical, since we have focused only on the number density evolution of galaxies as characterized by their baryonic properties, which are subject to influence from poorly constrained and crudely modeled sub-grid prescriptions for many physical processes.
To address this point we consider the number density evolution of the dark matter haloes directly, since they are relatively insensitive to the baryonic models.  
We select four galaxy populations using the same number density criteria that were employed in Figures~\ref{fig:multi_mass_nd} and~\ref{fig:multi_vd_evo}.
The result of tracking these four galaxy populations backward in time is shown in Figure~\ref{fig:multi_dm_evo}.
We find that the characteristic evolution of dark matter halo masses is fairly different from what was found for the stellar mass evolution -- especially at the high-mass end.
At the high-mass end, massive galaxies tend to quench owing to the AGN feedback prescriptions that have been implemented in our simulation.
This leads to relatively flat late-time growth rates for the stellar masses as shown in Figure~\ref{fig:multi_mass_nd}.
In contrast, no such late-time flattening of the halo mass growth rate is present in Figure~\ref{fig:multi_dm_evo}.
All haloes continue to grow rapidly until the present day.

We next consider what this means for the number density evolution of this galaxy population as shown in the right panel of Figure~\ref{fig:multi_dm_evo}.
The black dashed lines indicate the number density evolution calculated in Section~\ref{sec:Results_CMF} based on galaxy stellar mass and are therefore identical to those presented in Figures~\ref{fig:multi_mass_nd} and~\ref{fig:multi_vd_evo}.
We find that -- despite the visible differences in the stellar mass and halo mass growth trajectories -- the median number density evolution is nearly identical regardless of whether we use stellar mass, stellar velocity dispersion, or dark matter halo mass to trace the number density of galaxies in time.
This allows us to conclude that the implementation of baryonic physics and feedback processes as included in our simulations does not dominate the number density evolution of galaxies.
Rather, the stochastic growth rate of the underlying dark matter halo is the primary driver of the galaxy number density evolution that we find in our simulations.
We have repeated this analysis tracking galaxies forward in time, and have arrived at the same conclusion.
For this reason, we consider the derived median forward and backward number density evolution trends presented in this paper to be robustly tied to the underlying dark matter halo growth rates and to be relatively independent of the specific implementation of baryon physics/feedback adopted in our simulation.

Nevertheless, we still caution that some of the galaxy properties considered in this paper are subject to influence from the adopted physics/feedback prescriptions employed in our simulations.
For example, the size-mass relation derived for the Illustris galaxy population is shifted to larger sizes for low mass galaxies when compared against observations.
This is likely an indication of a shortcoming in either our treatment of the ISM equation of state or feedback implementation and could impact the derived velocity dispersion explored in Section~\ref{sec:Results_VDF}.
Similarly, the galaxy stellar mass function obtained within the Illustris simulation broadly agrees with observations across a wide range of redshifts~\citep{Torrey2014, Genel2014, Sparre2014, Somerville2014}, but differs in detail.
It is therefore possible that future generations of simulations or semi-analytic models that better match the galaxy mass distribution consistent with observations could yield somewhat different median or scattered comoving number density evolution rates.
Although we feel confident that the non-constant comoving number density fits prescribed in this paper are an improvement over the constant comoving number density assumption commonly applied in the literature, the previously mentioned caveats along with those discussed in~\citet{Nelson2015} should be kept in mind when applying the evolving cumulative number density fits presented in this paper.

\subsection{Additional Parameter Dependencies}
A wide range of galaxy properties beyond the stellar mass, velocity dispersion, and dark matter halo mass are tracked in our simulations.
We can therefore consider the role that several other galaxy parameters may play in predicting the scatter seen in the galaxy number-density evolution.
For example, it is reasonable to suspect that the relative late/early formation times of galaxies can be distinguished based on galaxy color.
Given the abundance of basic information we have about an observable galaxy population at some redshift (e.g., $z=0$ galaxy masses, sizes, star formation rates, colors, etc.), how deterministically can we predict an individual galaxy's evolutionary history?
We have shown in this paper that galaxy populations of similar stellar mass will have large scatter in their formation histories, and it is not immediately clear to what extent we can differentiate between galaxies that will grow faster or slower compared to their peers of similar initial mass by considering additional galaxy properties.

We adopt the most straightforward method to identify additional parameter dependencies that follows the same approach used throughout this paper.
Specifically, we perform an ordinary linear regression using the redshift $z=0$ galaxy stellar masses, stellar velocity dispersions, sizes, star formation rates, and $g-r$ galaxy colors.
We adopt the stellar half-mass radii as a proxy for galaxy size and the $g-r$ color is calculated based on the~\citet{bc03} stellar photometric catalogs as tabulated in~\citet{Torrey2015}.
The fit that we apply takes the general form 
\begin{equation}
{\rm Log}_{10}(N) = \sum_i \sum_{j=0}^4 C_{i,j}   x_i ^j 
\label{eqn:all_params_fit}
\end{equation}
where $i$ is a summation over galaxy properties ($i=0$ is stellar mass, $i=1$ is stellar velocity dispersion, etc.), $j$ is a summation over polynomial expansion order, and $C_{i,j} = c_{0,i,j} + c_{1,i,j}  z + c_{2,i,j}  z ^2$ are the redshift-dependent coefficients.
The regression is performed using the tracked number density of the galaxy population with redshift $N=N(z)$ as well as the redshift $z=0$ galaxy properties.
The regression yields the best possible fit to the number-density evolution of the galaxy population backward in time based on the properties that are known at redshift $z=0$.
If there are residual correlations driving the scatter seen in the number-density evolution in Figures~\ref{fig:multi_mass_nd} and~\ref{fig:multi_vd_evo}, then they will be captured with this fitting procedure.
We note that the fourth order polynomial in Equation~(\ref{eqn:all_params_fit}) gives fits to the number density evolution which are equally good (i.e. errors of order a few percent) as those given in Section~\ref{sec:Results_CMF} when stellar mass is the only parameter considered, which makes this a fair comparison.

\begin{figure}
\centerline{\vbox{\hbox{
\includegraphics[width=3.5in]{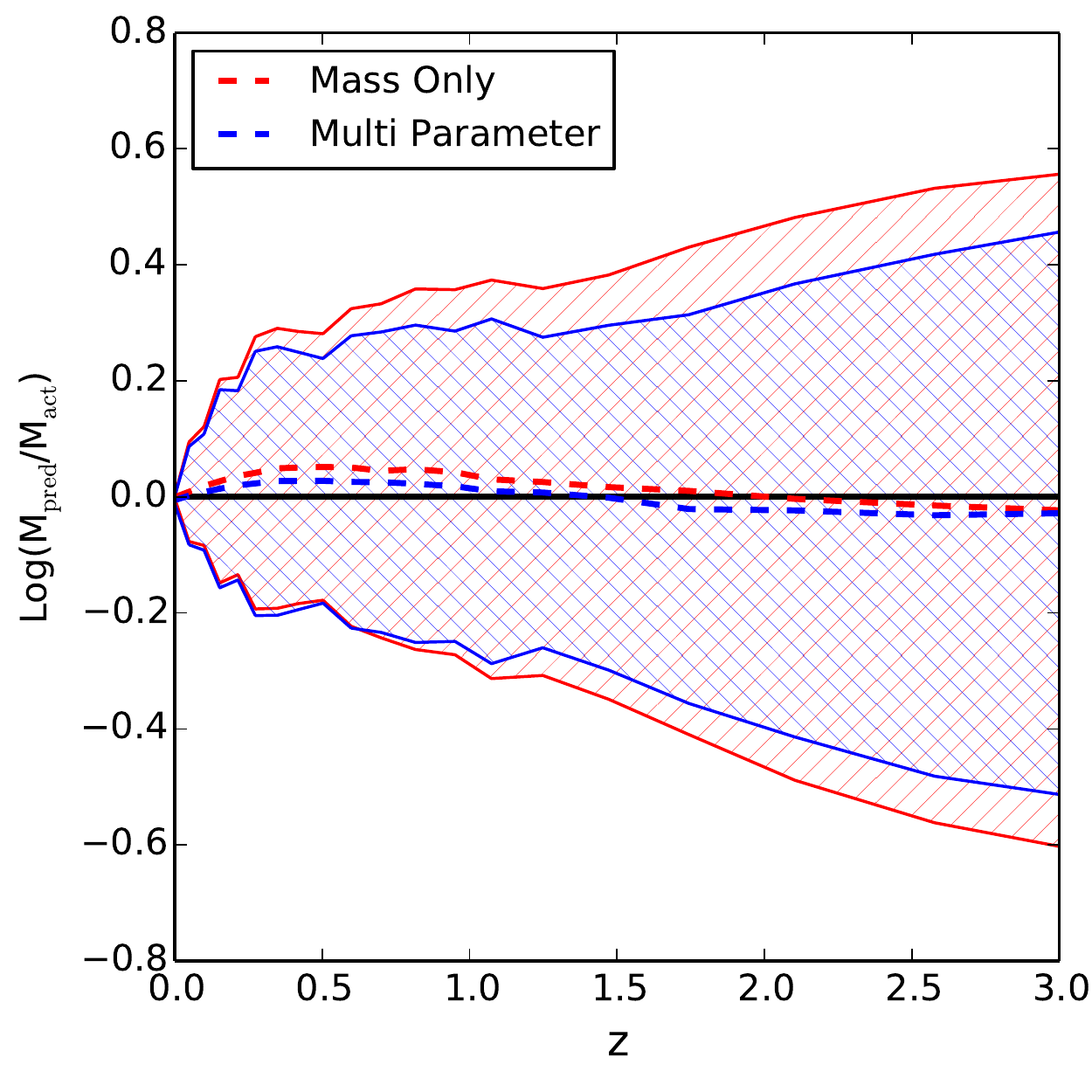}}}}
\caption{The median and $1\sigma$ standard deviation for the log ratio of the predicted mass to the actual tracked stellar mass (from the merger tree) as a function of time for $z=0$ Milky Way mass galaxies.  
Red lines indicate the predictions when $z=0$ stellar mass is the only parameter considered, and blue lines indicate the predictions when the redshift $z=0$ stellar velocity dispersions, stellar half mass radii, star formation rates, and $g-r$ galaxy colors are also included.
The scatter is reduced by only $\sim20\%$ when these additional parameters are considered.  }
\label{fig:full_fit_scatter}
\end{figure}

Given the number-density evolution fit in Equation~(\ref{eqn:all_params_fit}), we can derive the stellar mass evolution using the tabulated CMF given in Section~\ref{sec:Results_CMF}.
We are then able to quantify the reduction in the scatter of this fit by considering the error in the resulting stellar mass estimates.
Figure~\ref{fig:full_fit_scatter} shows the median and standard deviation of the log ratio of the predicted mass to the actual mass at several redshifts for the Milky Way mass selected galaxies using both the mass only fits given in Section~\ref{sec:Results_CMF}, as well as the multi-parameter fit given in Equation~(\ref{eqn:all_params_fit}).
We find that the multi-parameter fits provide a median error which is similar to the mass-only fit, but that the one-sigma standard deviation in the scatter is reduced by $\sim0.1$ dex.
While the multi-parameter fit is an improvement over the ``mass only'' fit, this amounts only to a $\sim20$\% reduction in the scatter.  
Even with an accounting of the galaxy stellar masses, sizes, star formation rates, colors, and stellar velocity dispersions entering into our analysis, 
our improved fit still has a 0.3(0.4) dex standard deviation by redshift $z=2(3)$.

The lack of significantly reduced scatter indicates that a direct and unambiguous linking cannot be achieved between high and low redshift populations given the {\it simulated} galaxy stellar masses, sizes, star formation rates, colors, and stellar velocity dispersions alone.
We do not rule out the possibility of being able to deterministically connect high redshift and low redshift galaxy populations in a direct progenitor-descendant link, but our results indicate that this would require information beyond the quantities explored here.
We have performed the same exercise tracing galaxies forward in time, and found similar results (i.e. a $\sim20\%$ reduction in scatter).
While marginally reduced, the scatter is still a significant component of the overall mass evolution.

We caution again that some of the galaxy properties considered in this section are subject to influence from the adopted physics/feedback prescriptions employed in our simulations.
We specifically note that although the simulated galaxy stellar mass function from Illustris broadly agrees with observations, the most massive galaxies continue to experience intermittent periods of star formation activity at late times that can lead to non-zero SFRs and greenish galaxy colors.
Both of these may adversely impact our ability to decompose mass-matched galaxy populations into late- and early-forming subsamples.
It will therefore be interesting to reconsider this problem using other currently available numerical simulations~\citep[e.g.,][]{Schaye2015} which employ different physical/feedback prescriptions~\citep{Crain2015} or with future generations of large volume galaxy formation simulations or semi-analytic models.

\subsection{Progenitor/Descendant Tracking Asymmetry}
\label{sec:tracking_asymmetry}
We have found in Section~\ref{ref:fow_back_comp} that tracing galaxies forward in time yields distinctly shallower inferred mass growth rates than tracing galaxies backward in time.
A similar manifestation of this effect is the qualitatively different number-density evolution for galaxies as they are traced  forward and backward in time.
Physically, tracing galaxies forward and backward captures different processes.
When tracing galaxies forward in time, a significant fraction of the tracked galaxy population can be ``lost" owing to merger events when the galaxy being tracked is swallowed by a more massive system.
The forward tracks therefore roughly capture the median mass evolution of the {\it surviving} galaxy population.
Tracing galaxies backward in time contains no analogous loss of systems owing to mergers.
By definition, any galaxy which exists in the simulation at redshift $z=0$ is a main branch.
The backward tracks therefore roughly capture the median mass evolution of the main progenitor galaxies.
Since these two tracking methods capture different physical galaxy populations, it should perhaps not surprise us that they yield qualitatively similar but quantitatively different mass evolution tracks.
However, if we select only main branches in both the forward and backward tracking analysis, we find that a nearly identical bias still persists between the inferred mass evolution in each direction.
The reason for this is that while the forward tracking does indeed suffer from a net reduction of tracked galaxies with time, 
the systems which are lost owing to mergers are more-or-less randomly sampled from the initial population (there are marginal correlations with environment, but these leave a non-detectable signal).

The main effect that drives the difference in forward-backward mass tracking is the asymmetric sampling of galaxy scattered growth given the initially selected galaxy population.
A population of galaxies selected at low redshift will naturally contain some subset of galaxies which had anomalously fast growth histories (i.e. which originated from much lower masses).
Although these anomalously fast growth histories only apply to a small fraction of the galaxy population, the steep nature of the galaxy stellar mass function implies a much higher abundance of low-mass galaxies that are able to follow these tracks.
Therefore, there is a conditional probability set by the shape of the high-redshift galaxy stellar mass function that tends to sample fast growth histories when tracing galaxies backward in time.
Tracking galaxies forward in time yields no similar conditional probability.
Instead, the primary source of galaxy dispersed growth histories is simply the scatter in the galaxy stellar mass function.
We leave a more formal exploration of the various mechanisms that drive galaxy number-density evolution to a future study.

\section{Conclusions}
\label{sec:Conclusions}
In this paper we have studied the stellar mass, central stellar velocity dispersion, dark matter halo mass, and corresponding comoving number-density evolution of galaxies using the Illustris hydrodynamical galaxy formation simulation.
We have compared the evolutionary paths of galaxy populations obtained by assuming that galaxies preserve their number density in time (the so-called constant number density ansatz) and by directly tracking the simulated galaxies backward and forward in time via the available merger trees.

Our main conclusions are as follows:  
\begin{itemize}
\item We provide a simple tabulated function that gives the cumulative stellar mass function (CMF) and cumulative stellar velocity dispersion function (CVDF)  from $z=0$ to $z=6$ in the Illustris Simulation (Equation~\ref{eqn:CumMassFunc} and Tables~\ref{table:CMF_fit} and~\ref{table:CVDF_fit}).  
This simple function can be used -- as we do in this paper -- to infer the stellar mass growth of galaxies at a fixed number density.
The cumulative stellar mass function found in the Illustris Simulation can be compared against observations, and we note that previous studies have presented such a comparison~\citep{Torrey2014, Genel2014} with favorable results.  
The functional fit provided in this paper for the differential and cumulative galaxy stellar mass function should help facilitate future comparisons with simulated data and semi-analytic results.
\item We trace galaxies forward and backward in time using merger trees from the Illustris simulations and find that galaxy populations do not evolve along constant comoving number-density tracks.
They fail to do so because of the combined influence of galaxy mergers and scattered galaxy growth rates.
We find that galaxies that are initially similar in their stellar mass, dark matter mass, or central stellar velocity dispersion diverge with time.
\item We find that the central stellar velocity dispersion evolves only mildly with redshift owing to the combined effects of mass and size growth.
Despite the mild velocity dispersion evolution we find that velocity dispersion yields a number density evolution that is not improved over that found for stellar mass or dark matter mass assigned number density evolution.
In fact, we find that the evolution of the number-density distribution of galaxies evolves nearly identically regardless of whether one uses  stellar mass, dark matter mass, or central stellar velocity dispersion to assign number density.
\item There is a systematic bias between the median mass growth rate inferred from constant comoving number-density analysis and merger tree analysis that we capture in our simulations.  
This bias is driven by a systematic evolution in the median number density of a galaxy population when traced in time.  
The median offset in stellar mass growth histories is only a factor of $\sim$2(4) when tracing Milky Way type galaxies out to redshift $z=2(3)$.  
However, we emphasize that this offset is systematic, and can be corrected for by accounting for the median number-density evolution of galaxies with time.
\item We provide a simple tabulated function that describes the number-density evolution of simulated galaxies both forward and backward in time (Equation~\ref{eqn:CumMassFunc} with Tables~\ref{table:Zevo_fit}-\ref{table:fwd3}).
We encourage the use of this simple form in place of the widely applied constant comoving number density.
While the non-constant comoving number-density evolution does {\it not} capture the scattered growth rates that are present for our simulated galaxy population, it does account for the first order offset for the median galaxy mass and number density evolution.
\item A fundamental asymmetry exists between progenitor and descendant tracking.
We find that the mass trajectories identified by following progenitor and descendant galaxy populations in time yield an offset of a factor of a few, which is systemically biased toward faster growth rates when tracing galaxies backward in time.
This implies that the progenitors of Milky Way or other mass galaxies would in fact be on average lower in mass than would be implied from a constant comoving number-density analysis.
This has direct implications for quoted, e.g., Milky Way mass progenitor mass evolutionary histories in the literature.
\item The scatter in the mass formation histories for any initially similar galaxy population is large.
We show that the simulated progenitors of present day Milky Way mass galaxies span at $z\sim2$ more than one order of magnitude in stellar masses.
We apply a regression including several galactic properties beyond stellar mass (size, star formation rate, galaxy color, and stellar velocity dispersion) and find that the error in the mass/number-density evolution can only be improved marginally (by $\sim$20\%).
\item We argue that the intrinsic scatter in galaxy growth rates implies that one cannot unambiguously identify galaxy progenitor/descendant populations between different observational epochs.  
\end{itemize}
In light of these conclusions, statistical methods for linking progenitor and descendant galaxy populations may be better suited for observationally deriving galaxy mass, size, and morphology evolution.

\section*{Acknowledgements} 

PT acknowledges support from NASA ATP Grant NNX14AH35G. 
SW is supported by the National Science Foundation Graduate Research Fellowship under grant number DGE1144152.
FM acknowledges support from the MIT UROP program.
RM acknowledges support from the DOE CSGF under grant number DE-FG02-97ER25308.
AP acknowledges support from the HST grant HST-AR-13897. 
Support for C-PM is provided in part by the Miller Institute for Basic Research in Science, University of California Berkeley. 
VS acknowledges support by the DFG Research Centre SFB-881 The Milky Way System through project A1, and by the European Research Council under ERC-StG EXAGAL-308037. 
LH acknowledges support from NASA grant NNX12AC67G and NSF grant AST-1312095.

The Illustris-1 simulation was run on the CURIE supercomputer  at  CEA/France  as  part  of  PRACE  project RA0844, and the SuperMUC computer at the Leibniz Computing  Centre,  Germany,  as  part  of  GCS-project  pr85je.
The further simulations were run on the Harvard Odyssey and CfA/ITC clusters, the Ranger and Stampede supercomputers at the Texas Advanced Computing Center through XSEDE,  and  the  Kraken  supercomputer  at  Oak  Rridge National Laboratory through XSEDE.
The analysis presented in this paper was conducted on the joint MIT-Harvard computing cluster supported by MKI and FAS.

\bibliographystyle{apj}


\appendix
\section{Non-cumulative Galaxy Stellar Mass Function}
\label{sec:DiffMassFunction}
\begin{figure}
\centerline{\vbox{\hbox{
\includegraphics[width=3.5in]{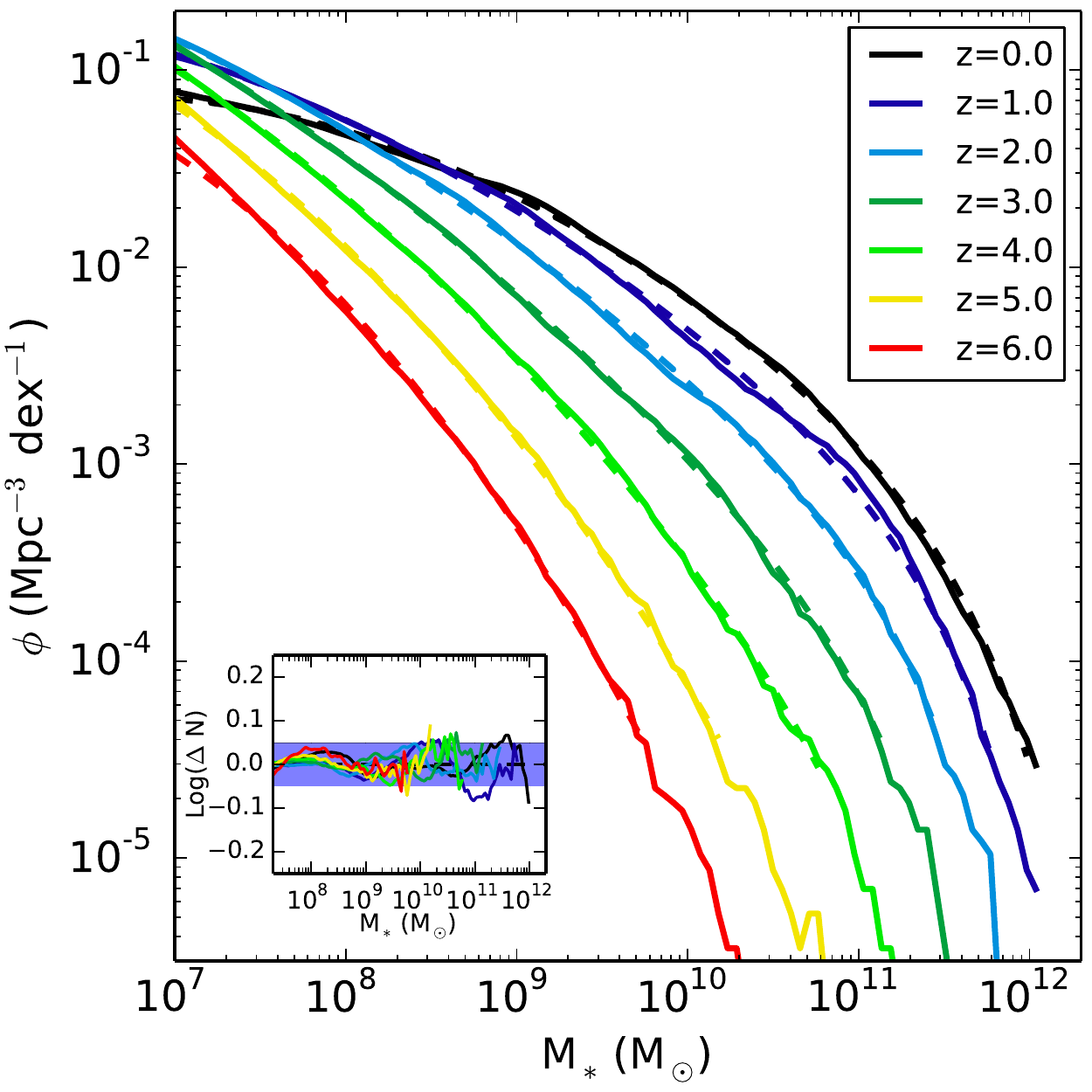}
}}}
\caption{Galaxy stellar mass functions derived from the galaxy populations formed in Illustris are shown at several redshifts as indicated in the legend.  
The dashed lines shown within indicate the galaxy stellar mass function fitting functions.
The fitting functions approximate the actual galaxy stellar mass function at all redshifts reasonably well, with the ``error" associated with these fits in the panel inset.
 }
\label{fig:diff_number_density}
\end{figure}

In Section~\ref{sec:CMF} we provided tabulated fits to the cumulative galaxy stellar mass function.  
Although useful for this paper, the CMF is less commonly used in the literature compared to the (differential) galaxy stellar mass function.
Here, we provide similar fits to the (differential) galaxy stellar mass function from the Illustris simulation that can be used easily for comparisons against other simulations or observational data sets.
We adopt a functional form of
\begin{equation}
\phi = \frac{dN}{d\mathrm{Log}M_*} =  A \; \tilde{ M}_* ^{\alpha + \beta \mathrm{Log}\tilde{ M}_*  } \; {\rm exp} ({-\tilde{ M}_* }) 
\label{eqn:DiffMassFunc}
\end{equation}
where $\tilde{ M}_* =M_*/(10^{\gamma} M_\odot)$ and the fit coefficients are allowed to vary with redshift as described in equations~\ref{eqn:A}-\ref{eqn:E}.
We identify the best fit coefficients using an ordinary regression based on the tabulated differential stellar mass function over the redshift range $0<z<6$.  
The galaxy stellar mass functions taken directly from the simulations and the associated best fits are shown in Figure~\ref{fig:diff_number_density} as solid and dashed lines respectively.
The inset shows the errors associated with the fits, which are marginally larger than what was found for the CMF.
However, the error remains well below 10\% for the full resolved redshift, mass, and number density.
The appropriate limits on this fitting function cover the mass range $10^7 M_\odot < M_* < 10^{12} M_\odot$, mass function values $\phi > 3 \times 10^{-5} \mathrm{Mpc}^{-3} \mathrm{dex}^{-1}$, and redshift range $0<z<6$.  
The best fit coefficients can be found in Table~\ref{table:DMF_fit} and a basic python script to evaluate the mass functions can be found online.\footnote{\url{https://github.com/ptorrey/torrey_cmf}}

\begin{table}
\begin{center}
\caption{The best fit parameters to the redshift-dependent differential mass function presented in Equation~(\ref{eqn:DiffMassFunc}) are given.}
\label{table:DMF_fit}
\begin{tabular}{ c  c c c c c  }
  &      \multicolumn{1}{|c|}{$i=0$}     &       \multicolumn{1}{|c|}{$i=1$}      &     \multicolumn{1}{|c|}{$i=2$}     \\
\hline 
\hline 
$a_i$ & -3.082270  & 0.091113  & -0.125720   \\
$\alpha_i$ & -0.675004  & 0.091193  & -0.049466   \\
$\beta_i$ & -0.043321  & 0.025282  & -0.007046   \\
$\gamma_i$ & 11.512307  & -0.190260  & 0.021313   \\
\hline
\hline
\end{tabular}
\end{center}
\end{table}

\end{document}